\documentclass [10pt,a4paper]{article}
\usepackage{jheppub}

\usepackage[latin1]{inputenc}
\usepackage{graphicx}
\usepackage{color}
\usepackage{amsfonts,amsthm,amssymb}
\usepackage{bm,bbm}
\usepackage[OT2,OT1]{fontenc}
\usepackage{nicefrac}
\usepackage{axodraw}

\usepackage{verbatim}
\usepackage{units}
\usepackage{mathrsfs}
\usepackage{amsmath}
\usepackage{amssymb}
\usepackage{esint}

\newcommand\cyr{%
\renewcommand\rmdefault{wncyr}%
\renewcommand\sfdefault{wncyss}%
\renewcommand\encodingdefault{OT2}%
\normalfont
\selectfont}
\DeclareTextFontCommand{\textcyr}{\cyr}

\def\beq{\begin{equation}}
\def\eeq{\end{equation}}
\newcommand{\be}{\begin{eqnarray}}
\newcommand{\ee}{\end{eqnarray}}

\renewcommand{\texttt}{{}}

\newcommand{\dal}{{\dot\alpha}}
\newcommand{\db}{{\dot\beta}}

\def\bs{\begin{subequations}}
\def\es{\end{subequations}}

\def\cD{\mathcal{D}}

\newcommand{\tia}[1]{}

\newcommand{\bea}{\begin{eqnarray}}
\newcommand{\eea}{\end{eqnarray}}
\newcommand{\beas}{\begin{eqnarray*}}
\newcommand{\eeas}{\end{eqnarray*}}
\newcommand{\bal}{\begin{aligned}}
\newcommand{\eal}{\end{aligned}}

\def\({\left(}
\def\){\right)}

\begin{document}

%\begin{frontmatter}

%\title{
%{\bf A UNIQUE FINITE QUANTUM GRAVITY %nique finite quantum gravity
%} %s beta functions
%}

\title{
{\bf  %Super-renormalizable Gravity}: \\
%Finite 
Classical and Quantum 
Nonlocal Supergravity}% in Superspace}  %s beta functions
}

\author[a,\hspace{0.6mm} b]{Stefano Giaccari} 
\author[a]{Leonardo Modesto}

\affiliation[a]{Department of Physics \& Center for Field Theory and Particle Physics, \\
Fudan University, 200433 Shanghai, China}

\affiliation[b]{Theoretical Physics Division of Particles and Fields\\
                    Faculty of Science, University of Zagreb\\
                    Bijeni\v{c}ka 32, HR-10000 Zagreb, Croatia\\}
                    
\emailAdd{sgiaccari@phy.hr}
\emailAdd{lmodesto@fudan.edu.cn}

%\date{\small\today}

\abstract{
We derive the $N=1$ supersymmetric extension for a class of weakly nonlocal four dimensional gravitational theories.
The construction is explicitly done in the superspace where the off-shell supersymmetry is manifest. The tree-level perturbative unitarity is therefore explicitly proved both in superfield formalism and in field components. 
For the minimal nonlocal supergravity the spectrum is the same as in the local theory and in particular 
it is ghost-free. 
The supersymmetric extension of the super-renormalizable Starobinsky theory 
and of two alternative massive nonlocal supergravities are found as straightforward applications of the formalism. 
Power-counting arguments ensure super-renormalizability with milder requirement for the asymptotic behavior of form factors than in ordinary nonlocal gravity. The most noteworthy result, common to ordinary supergravity, is the absence of quantum corrections to the cosmological constant in any regularization procedure. We cannot exclude the usual one-loop quadratic divergences. However, 
local vertices in the superfields, not undergoing renormalization, can be introduced to cancel out such divergences, thus playing the role of ``super-killer" operators. %terms for the divergences,
Therefore, quantum finiteness is certainly achieved in dimensional regularization and most likely also 
in the cut-off regularization scheme.
We also discuss the $n$-point scattering amplitudes making use of a general field redefinition theorem implemented in the superspace. Finally, we show that all the exact solutions of the local supergravity in vacuum are solutions of the nonlocal one too. In particular, we have the usual Schwarzschild singularity despite the presence of matter, contrary to the expectation that it should automatically be smeared out by the nonlocal form factor. 
We infer that weak nonlocality, even in the presence of minimal supersymmetry,  is not sufficient to solve the spacetime singularities issue, although the theory is finite at quantum level. }

%Extended $N=2$ supergravity, $N=4$ or $N=8$, might be free from divergences 
%also at one loop. The extended supergravities  would then result finite at any order in the loop expansion.

\maketitle
%\tableofcontents
%\keywords{Nonlocal Field Theory, Higher Derivative Gravity, Quantum Gravity}

%\maketitle

%\PACS 04.60.-m, 11.10.Lm, 11.25.Yb

% 04.60.-m	Quantum gravity
% 11.10.Lm	Nonlinear or nonlocal theories and models
% 11.25.Yb	M theory

\date{\small April 7, 2014}

%\tableofcontents

\section{Introduction}

Higher-derivative gravity has been the object of enduring interest in theoretical physics research due to its relevance to a quite wide range of still debated crucial issues. It naturally comes on the scene as the  straightforward consequence of the cancellation %renormalization 
of divergences in quantum field theories in curved background. Moreover, the introduction of terms quadratic in the curvature leads to renormalizable theories of quantum gravity, but the price to pay for renormalizability is the introduction of unphysical ghost modes in the theory. In $D=4$ the theory described by the following action% an action of the kind
\footnote{We remind that in $D=4$ a linear combination of the curvature-square $ \mathscr{R}^{2}$, Ricci-square $\mathscr{R}_{\mu\nu}\mathscr{R}^{\mu\nu}$ and Weyl-square 
$\mathscr{C}_{\mu\nu\rho\sigma} \mathscr{C}^{\mu\nu\rho\sigma}$ is a total derivative by the Gauss-Bonnet theorem. }
\begin{equation}
S_{\rm Stelle} = \frac{2}{\kappa^{2}}\int d^{4}x\sqrt{-g}\left(\mathscr{R}-2\Lambda_{\rm Cosmo}+\gamma_0 \mathscr{R}^{2}+ \gamma_2 \mathscr{R}_{\mu\nu}\mathscr{R}^{\mu\nu}
%+ \gamma_w \mathscr{C}_{\mu\nu\rho\sigma} \mathscr{C}^{\mu\nu\rho\sigma}
\right)\,   \label{eq:StelleQuad}
\end{equation}
is indeed power-counting renormalizable \cite{Stelle}, but the spectrum %(at linearized level)
contains (besides the usual massless graviton) a massive spin $2$ poltergeist (negative-metric state)
and a positive-metric state scalar field. In the limit $\gamma_2\rightarrow0$ renormalizability is lost, but the poltergeist decouples and we are left with a totally acceptable spectrum, i.e. the graviton plus a physical scalar (provided $\gamma_0 > 0$, otherwise it is a tachyon). This is indeed the Starobinsky's celebrated model compatible with self-inflation \cite{Starobinsky:1980te}. 

The action \eqref{eq:StelleQuad} is just one example in a general class of local and nonlocal theories 
at most quadratic in the curvature tensor. This range of theories can be explicitly written making use of the following short notation, 
\begin{equation}
\hspace{-0.1cm}
S_{\rm NL} =
\frac{2}{\kappa^{2}}\int \!\! d^{4}x\sqrt{-g}\left(\mathscr{R}-2\Lambda_{\rm Cosmo}+ \mathscr{R}\gamma_0(\square_{\Lambda})\mathscr{R}
+\mathscr{R}_{\mu\nu} \gamma_2(\square_{\Lambda} )\mathscr{R}^{\mu\nu} 
+  \mathscr{C}_{\mu\nu\rho\sigma} \gamma_w (\Box_\Lambda)  \mathscr{C}^{\mu\nu\rho\sigma}
\label{NL}
\right) , 
\end{equation}
where $\gamma_0\left(\square_{\Lambda}\right)$, $\gamma_2\left(\square_{\Lambda}\right)$,
and $\gamma_w\left(\square_{\Lambda}\right)$
are functions of the d'Alembertian operator $\square_{\Lambda}= \square/\Lambda^{2}$ with $\Lambda$ an arbitrary mass scale. The asymptotic UV behavior of such functions can always be chosen to make the theory super-renormalizable with divergences showing up only at one loop level in the perturbative expansion \cite{shapiro3, shapiroComplex, GBV, Shapirobook}. Moreover, the infrared limit is constrained by the requirement that the usual Einstein gravity is recovered at low energies  \cite{Maggiore,TsujikawaModestoIR,Deser, Odintsov, Barvinsky}. Actually, it has been explicitly shown \cite{Kuzmin, Tombo, Krasnikov, Khoury:2006fg, modesto,modestoLeslaw, Briscese:2013lna} that entire functions can be chosen for the form factors in \eqref{NL} such that the weakly nonlocal theories of gravity attained in this way have the following fundamental characteristics:
(i) general covariance; 
(ii) weak nonlocality (or quasi-polynomiality) \cite{Efimov, Cnl1, Anselmi:2015dsa}; 
(iii) unitarity (ghost-freedom); 
(iv) super-renormalizability at quantum level. It is interesting to notice that such theories can actually be defined in any dimension $D$ with the simplification that in odd dimension there are no one loop divergences in dimensional regularization scheme \cite{ModestoFinite} and the theory turns out to be completely finite. UV finiteness  is 
achievable for any $D$ by introducing a local curvature potential $O(\mathscr{R}^3)$ that, while not modifying the UV behaviour of the graviton propagator, contributes to vertices whose couplings enter in the beta functions linearly. This allows to fix all the beta functions to zero by a suitable choice of a few couplings in front of the so called {\em killer operators} appearing in the potential \cite{modestoLeslaw}.
This is not a fine tuning because the result is one loop exact. 
The coupling of these theories to matter and gauge bosons has been also extensively studied in
\cite{ModestoMatter1, ModestoMatter2, unification2, BrisceseScalar}, 
producing evidence for a super-renormalizable or finite theory of all fundamental interactions
\cite{ModestoMatter1}. This is explicitly proved in \cite{ModestoMatter2}. 

At classical level the gravitational potential and some approximate black hole solutions turn out to be 
%there are evidences %has been found 
%supporting that 
%these theories can be considered  
``{\em singularity-free}" for the case of any physical matter satisfying the energy conditions  
\cite{M3, M4, ModestoMoffatNico,BambiMalaModesto2, BambiMalaModesto,calcagnimodesto, koshe1, BM, ModestoGhosts}. 
However, it has been recently showed in \cite{Yaodong}, and also confirmed in the Palatini formulation 
of the theory \cite{BriscesePalatini}, that all Einstein spacetimes (including Schwarzschild and Kerr) and the FRW universes, when gravity is coupled to conformal matter, are exact solutions. 
Therefore, {\em spacetime singularities are still present in nonlocal gravitational theories}.  
%It is the delocalization mechanism here at work together with scale/conformal invariance to make the singularities disappearance. 

In this paper, we deal with the supergravity embedding of these weakly nonlocal theories of gravity. There are two reasons that make us to think this is a desirable goal to achieve. 

The $N=1$ higher-derivative supergravity was extensively studied in the past as a low-energy effective theory in the context of more fundamental theories of gravity, such us string theory, where very likely 
%it is almost unavoidable that 
%some amount of 
supersymmetry 
%is almost unavoidable 
plays a crucial role at some intermediate energy scale. 
Indeed, string theory contains an infinite tower of massive modes that, upon integrating out, require that the effective action for the massless modes cannot just be the standard two-derivatives supergravity action. On top of this, the Green-Schwarz mechanism for the cancellation of gauge and gravitational anomalies involves the introduction of suitable higher derivative terms.  Of course these higher derivative terms should be consistent with the absence of poltergeists in the spectrum, which led several authors to conjecture that the effective higher derivative action for the massless modes could only contain the super Gauss-Bonnet combination at the fourth order in the derivatives. 
The latter can always be {\em assumed} if we are just interested in the tree-level scattering amplitudes of on-shell massless particles that depend only on the coefficient of the square of the Weyl tensor for $D>4$.
Such interesting result about the S-matrix has recently been extended to actions of the kind \eqref{NL}, confirming that ``on the Einstein shell" the ghosts never show up for any effective action \eqref{NL} and the compatibility between the ghost-freedom of the string and its effective gravity is guaranteed a fortiori on shell.  Nevertheless, if we take seriously the full particle content of the effective gravity theory emerging from string theory, we should actually discuss the issue of unitarity in the general setting of the supersymmetric embedding of the action \eqref{NL}. 
This is tantamount to requiring that unitarity is achieved not just by a perturbative redefinition of the gravity fields carried out order by order in the derivative expansion, but non-perturbatively in the energy.

% just what we are going to do in this paper.

The second motivation is related to the delicate issue of UV finiteness. Supergravity theories naturally emerge as the zero slope limit of string theory, which solves the ultraviolet problem of quantum gravity a fortiori by assuming the existence of extended objects as the fundamental excitations. Nevertheless, the dimensional character of the Newton constant and the related nonlinearity of the action seem to point to the non-renormalizability of gravity and supergravity theories. However, local supersymmetry has been proven to play  a crucial role in delaying the onset of ultraviolet divergences. In particular, since the 1980's it has been known that the presence of counterterms in the maximal supergravity (and maximal super Yang-Mills theory) is constrained by non-renormalization theorems, which are analogous to the ones found for globally supersymmetric theories, leading to the general expectation that UV-divergences should 
show up at $3$-loops order in the four dimensional $N=8$ supergravity. However, explicit computations of supposedly divergent diagrams by improved unitarity method have shown perturbative finiteness at $3$- and $4$-loops orders, leading to a detailed reconsideration of the non-renormalization theorems. 
Indeed, previous results were based either on the construction of on-shell supersymmetric invariants or on off-shell formalism for the linearized theory only. The unexpected cancellations of divergences have thus found an explanation in terms of a more careful analysis of the full local supersymmetric Ward identities with the requirement of continuous duality symmetry and additional predictions have been done for cancellations at $5$- and $6$-loops orders. The same predictions for the onset of supergravity divergences can be obtained from a superstring perspective as well, but it is still mysterious how the UV finiteness of string theory can be inherited by the maximally supersymmetric effective field theory describing its low energy physics (see also the discussion in \cite{calcagnimodesto}.) 
In this sense $N=1$ weakly nonlocal supergravity is an interesting intermediate case where the interplay between the improved UV behaviour brought about by some amount of supersymmetry and the UV finiteness determined by the emergence of a fundamental scale can be explicitly investigated.

%The paper is organized as follows.
%In section

Finally, we believe that the theory here presented is an attractive field theory 
 proposal  for a super-renormalizable or finite unified theory of gravity and matter. 
 Of course, for this achievement we need 
 more supersymmetry or a multidimensional supergravity \cite{Mtheory, Acharya}, 
and it is well known that super-string theory is the best candidate 
 for a consistent unification of all fundamental interactions. 
 However, the off-shell power of quantum field theory is still the big lack of string theory. 
 With this paper we would revive supergravity as a fundamental theory after many years it was
 confined to the role of an effective field theory for string theory. 
 For people believing in supergravity we think this paper will be an enjoyable reading.

\section{$N=1$ Nonlocal Supergravity}
In this section we explicitly construct %the remind Einstein 
the weakly nonlocal supergravity theory 
in the Wess-Zumino $N =1$ superspace. 
%to establish the formalism used to construct local or nonlocal higher derivative theories. 
%We here use t
The formalism used in this section %that the reader 
may be found in standard textbooks (for instance \cite{Gates1, Buchbinder, WessBagger}). %Other basic formulas can be found in the appendix. 
In particular,  we shall adopt the conventions and definitions of \cite{Buchbinder}.

The Wess-Zumino superspace formulation involves the covariant derivatives $\mathcal{D}_A=\left(\mathcal{D}_a,\mathcal{D}_{\alpha}, \overline{\mathcal{D}}^{\dot\alpha}\right)$, satisfying the following minimal algebra, 
\be
&&\left\{\mathcal{D}_\alpha,\overline{\mathcal{D}}_{\dot\alpha}\right\}=-2i\mathcal{D}_{\alpha\dot\alpha}\,,
\quad 
%&&
\left\{\mathcal{D}_\alpha,\mathcal{D}_\beta\right\}=-4\overline{R}M_{\alpha\beta}\,,\quad \left\{\overline{\mathcal{D}}_{\dal},\overline{\mathcal{D}}_\db\right\}=4{R}\overline{M}_{\dal\db}\,, \\
&&\left[\mathcal{D}_{\alpha} ,  \mathcal{D}_{\beta\db}\right]=i\epsilon_{\alpha\beta}\left(\overline{R}\overline{\cD}_\db+G^\gamma{}_\dal\cD_\gamma\right)+\overline{\cD}_\db \overline{R}M_{\alpha\beta}\label{eq:Superalgebra} %\\
-i\epsilon_{\alpha\beta}\cD^\gamma G^{\delta}{}_\dal M_{\gamma\delta}+2i\epsilon_{\alpha\beta}\overline{W}_\db{}^{\dot\gamma\dot\delta}\overline{M}_{\dot\gamma\dot\delta}\,,\nonumber
\ee 
where the following torsion superfields appear: the chiral fields $W_{\alpha \beta \gamma}$ and ${R}$ with superspins $3/2$ and $0$ respectively, and 
a real axial vector $G_{\alpha\dot\alpha}$.  They satisfy the following Bianchi identities, 
\begin{eqnarray}
&&\overline{G}_{\alpha\dot\alpha}=G_{\alpha\dot\alpha}\,,\quad W_{\alpha\beta\gamma}=W_{\left(\alpha\beta\gamma\right)}\,, \quad 
%\nonumber\\
%&&
\overline{\mathcal{D}}_{\dot\alpha}R=0\,,\quad \overline{\mathcal{D}}_{\dot\alpha}W_{\alpha\beta\gamma}=0\,,\nonumber\\
&&\mathcal{D}_{\alpha}R=\overline{\mathcal{D}}^{\dot\alpha}G_{\alpha\dot\alpha} \,, \, 
\quad 
\label{eq:NLBianchi}
%\\
%&&
\mathcal{D}^\gamma W_{\alpha\beta\gamma}=\frac{i}{2}\mathcal{D}_{\alpha}{}^{\dot\alpha}G_{\beta\dot\alpha}+\frac{i}{2}\mathcal{D}_\beta{}^{\dot\alpha}G_{\alpha\dot\alpha}\,.
\end{eqnarray}

As can be seen by a direct inspection of their components,  $R$, $G_{\alpha\dot\alpha}$ and  $W_{\alpha \beta \gamma}$
are the superspace analogs of $\mathscr{R}$, $\mathscr{R}_{\mu\nu}$ and $ \mathscr{C}_{\mu\nu\rho\sigma}$ appearing in the previous section (\ref{NL}). 
%(there is here no reason to be confused by the notation here introduced because we will never use again purely Bosonic tensors.) 
All the local supersymmetric invariant operators can be built out of 
${R}$, $G_{\alpha\dot\alpha}$, $W_{\alpha \beta \gamma}$, 
and the covariant derivative in superspace $\mathcal{D}_A$.
%{\color{red}($x^A: x^\mu, \theta^\alpha, \theta^{\dot{\alpha}}$)}
%As for the Bosonic case we have the superspace version of the Gauss-Bonnet theorem saying that
%the following integrals sum to a topological invariant,
%\be
%\int d^8 z E^{-1} \left(  2 R \bar{ R }+ G^\mu G_{\mu} \right)  + \int d^8 z \hat{\phi}^3 
%W_{\alpha \beta \gamma} W^{\alpha \beta \gamma} \, . 
%\label{GB0}
%\ee
Through the introduction of the supervielbein $E_A=E_A{}^M\partial_M$ we also have a natural notion of invariant superspace integration $\int d^8z\,E^{-1} \mathcal{L}$  for a scalar superfield $\mathcal{L}$, where $E\equiv {\rm Ber} (E_A{}^M)\neq 0$ and Ber$(..)$ is the determinant in the superspace.  %We have t
The rule to integrate by parts is: %, namely 
\be
\int d^8z\,E^{-1} \mathcal{D}_A V^A=0\,. 
\ee
%which implies, i
In particular, if $\psi_\alpha(z)$ and $V_a(z)$ are spinor and vector superfields then, under proper boundary conditions, we have:
\begin{equation}
\int d^8z\,E^{-1} \mathcal{D}^\alpha \psi_\alpha=0, \qquad\int d^8z\,E^{-1} \mathcal{D}_a V^a=0\,.
\label{eq:Intparts}
\end{equation}
{We also have an invariant chiral integral $\int d^6z\, \hat{\varphi}^3\hat{\mathcal{L}}_c$, where 
$\mathcal{L}_c$ is a covariantly chiral scalar superfield ($\overline{\mathcal{D}}_{\dot\alpha} \mathcal{L}_c=0$) and $\varphi$ is the flat chiral compensator superfield of Einstein supergravity ($\overline{D}_{\dot\alpha}\varphi$) = 0, while $\hat{\varphi}$ will be shortly defined in (\ref{hatvar}). 
The chiral integral is defined in the so called chiral representation, in which every superfield $V$ is changed to}
\begin{equation}
\tilde{V}=e^{-\overline{W}}V\,,
\end{equation} 
where $W$ is one of the prepotential in terms of which the supergravity constraints can be solved (see \cite{Buchbinder} for their definition.) In particular we have
\be
\mathcal{L}_c=e^{\overline{W}}\hat{\mathcal{L}}_c\,,\quad {\overline{\partial}}_{\dot\mu}\hat{\mathcal{L}}_c=0 \,, \quad 
\varphi=e^{\overline{W}}\hat{\varphi}\,,\quad {\overline{\partial}}_{\dot\mu}\hat{\varphi}=0\,.
\label{hatvar}
\ee 
The chiral integration formula
\begin{equation}
\int d^6z\, \hat{\varphi}^3\hat{\mathcal{L}}_c=\int d^8z\,\frac{E^{-1}}{R}\mathcal{L}_c
\end{equation}
expresses the chiral integral as an ordinary integral in superspace. On the other hand, owing to the identity
\begin{equation}
\int d^8z\,E^{-1}\mathcal{L}=-\frac14\int d^8z\,\frac{E^{-1}}{R}\left(\overline{\mathcal{D}}^2-4R\right)\mathcal{L}\,,
\end{equation}
each integral over the superspace $\mathbb{R}^{4\vert4}$ can be reduced to a chiral-like integral. So the most general supergravity action should be a superfunctional of the kind
\begin{equation}
\frac1{\kappa^2}\int d^8z\,\frac{E^{-1}}{R}\mathcal{L}_c\left(T_{BC}{}^D,\mathcal{D}_AT_{BC}{}^D,\ldots\right)+{\rm h.c.}\,,
\label{eq:GeneralAction}
\end{equation}
where $\kappa$ is the gravitational coupling constant and $\mathcal{L}_c$ is a covariantly chiral scalar depending on the supertorsion and its covariant derivative up to an arbitrary order. The  mass dimensions of the quantities appearing in \eqref{eq:GeneralAction} are
\be
&&[ \kappa^{-2} ] = 2 \, , \quad [ d^8 z] = -2 \, , \quad [ E ] = 0 \, ,  \quad [ \mathcal{D}_\alpha ] =[ \overline{\mathcal{D}}_{\dot\alpha} ]= \frac12\,, \quad [ \mathcal{D}_a ] =1\,\\
&&  [ {R} ] =1 \, , \quad 
[ G_{\alpha\dot\alpha} ] = 1 \, , \quad [W_{\alpha \beta \gamma}] = 3/2 \, \nonumber. 
\label{Dimensioni1}
\ee
Actually, relevant examples of actions of the form \eqref{eq:GeneralAction} are
\begin{equation}
S_{SG}=-\frac{3}{\kappa^2}\int d^8z\,E^{-1}\,,
\label{eq:EinsteinSugra}
\end{equation}
which describes the old minimal Einstein supergravity with $12$ Bosonic components and $12$ Fermionic ones, and 
\be
\hspace{-0.7cm}
%\boxed{
S_{\rm QUAD} 
\! = \! \int \! d^8 z E^{-1} \left(-\frac{3}{\kappa^{2}}  
+\gamma_{{R}} \bar{R}  {{R}} 
+\gamma_{G} G^{\alpha\dot\alpha}    G_{\alpha\dot\alpha} 
%\right)  + \int \! d^8 z 
%+ \frac{E^{-1}}{ \mathcal{R}}  
+ \gamma_{W}{R}^{-1}
W_{\alpha \beta \gamma}  W^{\alpha \beta \gamma}  \right) ,  %}
\!\!\! 
\label{eq:SuperStelle}
\ee
which is just the supersymmetric version of \eqref{eq:StelleQuad}. Actually, in $D=4$ one of the three quadratic operators in \eqref{eq:SuperStelle} can be omitted because of the supersymmetric version of the well known Gauss-Bonnet theorem, stating that the superfunctional
\begin{equation}
\mathscr{P}=\int \! d^8 z E^{-1} \left({R}^{-1}
W_{\alpha \beta \gamma}  W^{\alpha \beta \gamma} 
-\frac12 G^{\alpha\dot\alpha} G_{\alpha\dot\alpha} 
+2\bar{R} {{R}} 
 \right)\,  %}
\!\!\! 
\label{SGB}
\end{equation}
is a topological invariant. Unfortunately, similarly to what happens for the theory described by the action \eqref{eq:StelleQuad}, a  theory containing terms quadratic in the gravitational curvatures propagates, besides the usual massless states, massive particles of spin $\leq 2$ \cite{Stelle,Whitt:1984pd}. It has been seen that also in the supersymmetric case, whereas the new massive particles with spin $\leq \frac12$ are physicaly acceptable positive-norm states (provided the quadratic terms are taken with the right sign, otherwise they bring about tachyons), the ones with spin $\geq 1$ are bound to be 
negative-norm states \cite{Ferrara:1978rk, Cecotti:1987sa, CFPS, Cecotti:1987mr}.

We want to adopt the same strategy that has been successfully used in the non supersymmetric case to write down nonlocal higher derivative actions which are ghost-free at perturbative level \cite{modesto,modestoLeslaw, Briscese:2013lna, Krasnikov, Tombo}. The first step is to write down the most general superaction  quadratic in the curvatures with an arbitrary number of derivatives. First of all, we can restrict our analysis to terms of mass dimension $2n$ ($n\in \mathbb{N}^+$)
\be
&& \int \! d^8 z E^{-1} R\, \mathcal{D}_\beta \ldots \overline{\mathcal{D}}_{\dot\beta}\ldots\overline{R}+
{\rm h.c.} \, ,\nonumber \\
&&\int \! d^8 z E^{-1} G^{\alpha\dot\alpha} \, \mathcal{D}_\beta\ldots\overline{\mathcal{D}}_{\dot\beta}
\ldots\, G_{\alpha\dot\alpha}+{\rm h.c.}\,,\label{eq:HigherDerivatives}\\
&&\int \! d^8 z \frac{E^{-1}}{R}W_{\alpha \beta \gamma} \,\mathcal{D}_\delta\ldots\overline{\mathcal{D}}_{\dot\delta}\ldots\, W^{\alpha \beta \gamma}+{\rm h.c.}\,,\nonumber
\ee
containing $4n$ spinorial derivatives, 
{the reason being that any other term of mass dimension $2n$ quadratic in the curvatures can be reduced to these by using the derivative algebra \eqref{eq:Superalgebra} and the Bianchi identities \eqref{eq:NLBianchi}}\footnote{Actually, also terms of the form
\begin{equation}
\int \! d^8 z {E^{-1}}\overline{R}\,\mathcal{D}_\delta\ldots\overline{\mathcal{D}}_{\dot\delta}\ldots\,\overline{R}+{\rm c.c.}
\end{equation}
can be introduced, but we shall ignore them in this context, because they are not expected to have a direct space-time analogue in the sense that they do not produce any pure vierbein-dependent contribution in components.
}.
Moreover, when we interchange the derivatives in \eqref{eq:HigherDerivatives}, we can assume they satisfy the usual flat superderivative algebra because the additional terms, due to the curved superspace algebra \eqref{eq:Superalgebra}, just give vertex terms with at least three curvatures. So we can restrict our attention to the following higher derivative terms, 
\be
&& \int \! d^8 z E^{-1} R\, \mathcal{D}^2 \overline{\mathcal{D}}^2 \ldots \mathcal{D}^2\overline{\mathcal{D}}^2\overline{R}\,,\nonumber\\
&& \int \! d^8 z E^{-1} G^{\alpha\dot\alpha} \, \mathcal{D}^2 \overline{\mathcal{D}}^2 \ldots \mathcal{D}^2\overline{\mathcal{D}}^2\, G_{\alpha\dot\alpha}\,,\label{eq:HigherDerivatives1}\\
&& \int \! d^8 z \frac{E^{-1}}{R}W_{\alpha \beta \gamma} \,\overline{\mathcal{D}}^2\mathcal{D}^2 \ldots\overline{\mathcal{D}}^2\mathcal{D}^2 \, W^{\alpha \beta \gamma}+{\rm h.c.}\,,\nonumber
\ee
with $2n$ derivatives $\cD$ and $2n$ derivatives $\overline{\cD}$. %because a
All the other terms of the kind outlined in \eqref{eq:HigherDerivatives} can be shaped into the form 
(\ref{eq:HigherDerivatives1}) 
up to purely vertex terms. In general we can write the terms of mass dimension $2n$ as
\be
&& \int \! d^8 z E^{-1} \, R(\Delta_R)^n\,\overline{R}+{\rm h.c.},\nonumber\\
&& \int \! d^8 z E^{-1} G^{\alpha\dot\alpha} \, ( \Delta_{G} )^n\, G_{\alpha\dot\alpha}+{\rm h.c.}\,,\label{eq:HigherDerivatives2} \\
&& \int \! d^8 z \frac{E^{-1}}{\overline{R}}\overline{W}_{\dot\alpha \dot\beta \dot\gamma} \,( \Delta_{W} )^n \, \overline{W}^{\dot\alpha \dot\beta \dot\gamma}+{\rm h.c.}\,,\nonumber
\ee
where $\Delta_R= \Delta_{G} = \Delta_{W} =\frac1{16}\mathcal{D}^2 \overline{\mathcal{D}}^2+\ldots$, and $\ldots$ stands for terms of dimension $2$ containing curvatures or derivatives of them. In particular a very natural choice is:
\begin{equation}
\Delta_R= \Delta_{G} = \Delta_{W} =\Delta_{-}=\frac1{16}\left(\mathcal{D}^2-4\overline{R}\right) \left(\overline{\mathcal{D}}^2-4R\right),
\end{equation}
with $\Delta_{-}$ the chiral projector, or another choice is:
\begin{equation}
 \Delta_{{R}}= \Delta_{G} = \Delta_{W}=\mathcal{D}^a\mathcal{D}_a\,.
\end{equation}
Making this last choice, we can assume the following form for the nonlocal supergravity action quadratic in the curvatures
\be
&& \hspace{-0.5cm}
%\boxed{
S_{\rm NL}  %\label{SNL}
%&&
\! = \! \int \! d^8 z E^{-1} \left(-\frac{3}{\kappa^{2}}  
+ {R} \, \gamma_{{R}}(\Delta_{{R}} ) \, \bar{{R}} 
+ G^{\alpha\dot\alpha} \,  \gamma_{G}( \Delta_{G} ) \, G_{\alpha\dot\alpha}\right.  
%\nonumber\\
%\right)  + \int \! d^8 z 
%+ \frac{E^{-1}}{ \mathcal{R}}  
+ \left. \frac12 {R}^{-1}
W_{\alpha \beta \gamma} \, \gamma_{W}( \Delta_{W} )\, W^{\alpha \beta \gamma} +{\rm h.c.} \right)   .  %}
\!\!\! 
\nonumber \\
&& \label{SNL}
\ee
This action is not only supersymmetric by construction, but also very compact and elegant. 
The first operator $-3/\kappa^{2} \int d^8 z E^{-1}$ is the Einstein-Hilbert supergravity action,
while the other operators are the superspace generalizations of the nonlocal operators quadratic in the curvatures. 

\section{Constructing the linearized supergravity}
\label{LinearSUGRA}
%{\color{red} In this section $\sigma$ is not defined at all.}
In order to address the problem of unitarity, we start by constructing a general %``old-minimal" \cite{}  
linearized theory for an higher derivative local or nonlocal $N=1$ supergravity theory. This is obtained generalizing the old minimal Einstein supergravity. 
Any such theory should contain the massless $\left(2,3/2\right)$ supermultiplet, describing both the graviton and gravitino degrees of freedom.  Such a supermultiplet is contained in the real axial vector superfield $H_{\alpha\dot\alpha}$, but appears together with other supermultiplets. % {\color{red} that we are going to introduce}.
The standard way to single out the different representations is through projectors, which sum to the identity \cite{Gates:2003cz}, namely\footnote{
Unlike the action (\ref{SNL}), after the rescaling $H^{\mu} \rightarrow \kappa H^{\mu}$
and $\sigma\rightarrow\kappa\sigma$ (which will be defined later in this section), one  find the following list of dimensions, 
\noindent
\be 
&& \left[x^{a}\right] = -1 \, , \quad   
\left[\partial_{a}\right] = 1 \, , \quad 
\left[\theta^{\alpha}\right] = \left[\bar{\theta}_{\dot{\alpha}}\right] = -1/2 \, , \quad 
\left[\partial_{\alpha}\right]=\left[\bar{\partial}^{\dot{\alpha}}\right] = 1/2 \, , \quad  
\left[d^{2}\theta\right]=\left[d^{2}\bar{\theta}\right]=1 \, , \quad  
\left[d^{4}x\right] = -4 \, , \quad \nonumber \\
&&
\left[d^{8}z\right]=-2 \,  , \quad  
\left[d^{6}z\right]=-3 \, , \quad  
%$\left[\mathcal{L}\right]=2$, $\left[\mathcal{L}_{c}\right]=3$, 
\left[H^{a}\right]=0 \, , \quad 
\left[\sigma\right]=1 \, , \quad  
\left[\mathcal{W}_{\alpha\beta\gamma}\right]=5/2 \, , \quad 
\left[\mathcal{R}\right]=2 \, , \quad 
\left[{\mathcal G} \right]= 2 \, ,  \quad 
\left[\sigma\right]=1 \, .
\ee
%\end{eqnarray*}
Notice that the curvature in this footnote and in the whole section are the linearized version of those in  (\ref{SNL}).
}
 
\begin{equation}
H_{\alpha\dot\alpha}=\left(\Pi_{0}^{L}+\Pi_{\nicefrac{1}{2}}^{L}+\Pi_{\nicefrac{1}{2}}^{T}+\Pi_{1}^{T}+\Pi_{\nicefrac{3}{2}}^{T}\right)H_{\alpha\dot\alpha}\,,
\end{equation}
where the projectors are defined by
\begin{eqnarray}
&& \Pi_{0}^{L}H_{\alpha\dot{\alpha}}  =  -\frac{1}{32}\square^{-2}\partial_{\alpha\dot{\alpha}}\left\{ D^{2},\overline{D}^{2}\right\} \partial^{\beta\dot{\beta}}H_{\beta\dot{\beta}}\,,\nonumber\\
&& \Pi_{\nicefrac{1}{2}}^{L}H_{\alpha\dot{\alpha}} = \frac{1}{16}\square^{-2}\partial_{\alpha\dot{\alpha}}D^{\gamma}\overline{D}^{2}D_{\gamma}\partial^{\beta\dot{\beta}}H_{\beta\dot{\beta}}\,,\nonumber\\
&& \Pi_{\nicefrac{1}{2}}^{T}H_{\alpha\dot{\alpha}}  =  \frac{1}{24}\square^{-2}\partial^{\beta}\phantom{}_{\dot{\alpha}}\left[D_{\beta}\overline{D}^{2}D^{\gamma}\partial_{\left(\alpha\right.}\phantom{}^{\dot{\beta}}H_{\left.\gamma\right)\dot{\beta}}+D_{\alpha}\overline{D}^{2}D^{\gamma}\partial_{\left(\beta\right.}\phantom{}^{\dot{\beta}}H_{\left.\gamma\right)\dot{\beta}}\right] ,\nonumber \\
&& \Pi_{1}^{T}H_{\alpha\dot{\alpha}}  =  \frac{1}{16}\square^{-2}\partial^{\beta}\phantom{}_{\dot{\alpha}}\left\{ D^{2},\overline{D}^{2}\right\} \partial_{\left(\alpha\right.}\phantom{}^{\dot{\beta}}H_{\left.\beta\right)\dot{\beta}}\,,\nonumber\\
&& \Pi_{\nicefrac{3}{2}}^{T}H_{\alpha\dot{\alpha}}  =  -\frac{1}{8}\square^{-2}\partial^{\beta}\phantom{}_{\dot{\alpha}}D^{\gamma}\overline{D}^{2}D_{\left(\gamma\right.}\partial_{\alpha}\phantom{}^{\dot{\beta}}H_{\left.\beta\right)\dot{\beta}}\,.\label{SuperProjectors}
\end{eqnarray}
Here the superscripts $L$ and $T$ denote longitudinal and transverse
projectors, while the subscripts $0$, $1/2$, $1$, $3/2$ stand
for superspin. 

The gauge invariance emerges whether only some of the above super-projectors are present in the action. It is the projector $\Pi_{\nicefrac{3}{2}}^{T}$ that singles out the superspin-$3/2$ part of $H_{\alpha\dot\alpha}$, and the corresponding projection $\Pi_{\nicefrac{3}{2}}^{T}H_{\alpha\dot{\alpha}}$ is invariant  under the following linearized gauge transformation, 
\begin{equation}
\delta H _{\alpha\dot{\alpha}}=\bar{D}_{\dot{\alpha}}L_{\alpha}-D_{\alpha}\bar{L}_{\dot{\alpha}}\label{eq:LinearGaugeTransformH} \, ,
\end{equation}
with $L_{\alpha}$ a unconstrained spinor superfield, which is characteristic of the linearized conformal supergravity. Indeed, the gauge freedom \eqref{eq:LinearGaugeTransformH} can be used to choose the Wess-Zumino gauge, namely 
\begin{equation}
H^{\alpha\dot{\alpha}}\left(x,\theta,\bar{\theta}\right)=\theta\sigma^{b}\bar{\theta}e_{b}\phantom{}^{\alpha\dot{\alpha}}\left(x\right)+i\bar{\theta}^{2}\theta^{\beta}\Psi^{\alpha\dot{\alpha}}\phantom{}_{\beta}\left(x\right)-i\theta^{2}\bar{\theta}_{\dot{\beta}}\bar{\Psi}^{\alpha\dot{\alpha}\dot{\beta}}\left(x\right)+\theta^{2}\bar{\theta}^{2}{A}^{\alpha\dot{\alpha}}\left(x\right)   .
\label{eq:WessZumino}
\end{equation}
This choice does not fix the gauge freedom completely, but actually carries about a representation for the Weyl transfomations: 
\begin{eqnarray}
\delta_{\rho}e_{b}\phantom{}^{a}  =  \rho \, e_{b}\phantom{}^{a}\,, \quad 
\delta_{\rho}\Psi^{a}{}_{\beta}  =  \frac{3}{2}\rho\Psi^{a}{}_{\beta}\,, \quad 
\delta_{\rho}\tilde{A}^{a}  =  2\rho\tilde{A}^{a}\,,
\end{eqnarray}
and the local chiral transformations:
\begin{eqnarray}
 \delta_{\Omega}e_{b}\phantom{}^{a}  =  0\,,\quad 
 \delta_{\Omega}\Psi^{a}{}_{\beta}  =  -\frac{i}{2}\Omega\Psi^{a}{}_{\beta}\,, \quad 
 \delta_{\Omega}\tilde{A}^{a}  =  \frac{1}{2}g^{ab}\partial_{b}\Omega\,,
\end{eqnarray}
for which $\tilde{A}^{a}$ is the corresponding gauge field. Furthermore, 
the spinor gauge parameter $\eta^{\alpha}\left(x\right)$ gives the
transformations
\begin{eqnarray}
\delta_{\eta}e_{b} \phantom{}^{a}  =  0\,, \quad 
\delta_{\eta}\Psi^{a}{}_{\beta} =  -i\left(\sigma^{b}\bar{\eta}\right)_{\beta}e_{b}\phantom{}^{a}\,, \quad 
\delta_{\eta}\tilde{A}^{a}  =  i\eta\Psi^{a}-i\overline{\eta\Psi^{a}}\,, 
\end{eqnarray}
which are called $S$-supersymmetry transformations. In a sense, $\Omega\left(x\right)$
and $\eta^{\alpha}\left(x\right)$ are supersymmetric partners of
$\rho\left(x\right)$. Therefore, we find that $H^{\alpha\dot{\alpha}}\left(x,\theta,\bar{\theta}\right)$  
(with such a gauge group) is a realization of conformal supergravity
rather than Einstein supergravity. In order to get the degrees of freedom that are expected 
for Einstein supergravity we have either
to put constraints on $L_{\alpha}$ to get rid of these additional symmetries,
or to introduce an additional superfield such that, even in the presence of the conformal group transformations, the theory still has the correct dynamical content .  The standard old minimal formulation of supergravity actually involves also a flat chiral field $\varphi$ satisfying $\bar{D}_{\dot{\alpha}}\varphi=0$, whose gauge transformation is given by
\begin{equation}
\delta\sigma=-\frac{1}{12}\bar{D}^{2}D^{\alpha}L_{\alpha}\,,
\label{eq:LinearGaugeTransformComp}
\end{equation}
where $\sigma$ is defined by $\varphi=e^{\sigma}$, $\bar{D}_{\dot{\alpha}}\sigma=0$.
For later convenience we define $\varphi=e^{iH_{0}}\hat{\varphi}e^{-iH_{0}}$ and
$\hat{\varphi}=e^{\hat{\sigma}}$, $\bar{\partial}_{\dot{\alpha}}\hat{\sigma}=0$.
In the weak field approximation we have $\varphi^{3}\approx1+3\sigma$ with 
\begin{eqnarray}
&& \hat{\sigma}  \approx  \frac{1}{6}h_{\alpha\dot{\alpha}}\phantom{}^{\alpha\dot{\alpha}}-\frac23i\theta\sigma_{a}\bar{\Psi}^{a}\left(x\right)+\frac13\theta^{2}B\left(x\right)
  =  \frac{1}{6}h_{\alpha\dot{\alpha}}\phantom{}^{\alpha\dot{\alpha}}-\frac23i\theta^{\alpha}\bar{\Psi}_{\alpha\dot{\alpha},}\phantom{}^{\dot{\alpha}}+\frac13\theta^{2}B\,,\nonumber\\
&& \bar{\hat{\sigma}}  \approx  \frac{1}{6}h_{\alpha\dot{\alpha}}\phantom{}^{\alpha\dot{\alpha}}-\frac23i\bar{\theta}_{\dot{\alpha}}\Psi^{\alpha\dot{\alpha},}\phantom{}_{\alpha}+\frac13\bar{\theta}^{2}\bar{B}\,, \quad \label{eq:WessZuminoComp} %\\
 \hat{\bar{\sigma}}  =  \left(e^{-2iH_{0}}\bar{\hat{\sigma}}\right)=e^{-2iH_{0}}\bar{\hat{\sigma}}e^{2iH_{0}}\,.\nonumber
\end{eqnarray}
We can conclude that the Einstein supergravity multiplet is given by a 
set of $12+12$ fields 
\be
\left\{ e_{m}\phantom{}^{a},\Psi_{m\alpha},\bar{\Psi}_{m}\phantom{}^{\dot{\alpha}},{A}_{m},{B},\bar{{B}}\right\}
\ee
that transform under spacetime general coordinate transformations, local Lorentz and
local supersymmetry transformations.
Therefore, we end up we the following remarkable result:
conformal supergravity and Einstein
supergravity can be written down as gauge theories with the same gauge
group, but with different dynamical content. 
Conformal supergravity
is described in terms of the gravitational superfield only, whereas
Einstein supergravity needs one more dynamical superfield, the chiral
compensator $\varphi$. %This implies i
In particular an action describing
the dynamical content of Einstein supergravity in terms of $H^{\alpha\dot{\alpha}}$
and $\varphi$ must be invariant under the gauge transformations \eqref{eq:LinearGaugeTransformH}
and \eqref{eq:LinearGaugeTransformComp} to have %, because this is necessary to have
the correct content of dynamical fields. 
This is indeed the case for the following linearized local Einstein supergravity action, 
\be
\hspace{-0.7cm}
S_{\rm SG}^{(2)} = \!\! \int \!\! d^{8}z\left\{ \frac{1}{8}H^{a}D^{\alpha}\bar{D}^{2}D_{\alpha}H_{a}-3\sigma\bar{\sigma}+\frac{1}{48}\left(\left[D_{\alpha},\bar{D}_{\dot{\alpha}}\right]H^{\alpha\dot{\alpha}}\right)^{2}-\left(\partial_{a}H^{a}\right)^{2}+2i\left(\sigma-\bar{\sigma}\right)\partial_{a}H^{a}\right\} \!\!   , \label{eq:LinearEinsteinSugra}
\ee
that, using the super-projectors \eqref{SuperProjectors}, turns in: 
\begin{equation}
S_{\rm SG}^{\left(2\right)}=\int d^{8}z\left\{ H^{\alpha\dot{\alpha}}\square\left(-\frac{1}{3}\Pi_{0}^{L}+\frac{1}{2}\Pi_{\nicefrac{3}{2}}^{T}\right)H_{\alpha\dot{\alpha}}-3\sigma\bar{\sigma}-i\left(\sigma-\bar{\sigma}\right)\partial_{\alpha\dot{\alpha}}H^{\alpha\dot{\alpha}}\right\} . \label{eq:ProjLinearEinsteinSugra}
\end{equation}
The gauge invariance is evident rewriting \eqref{eq:ProjLinearEinsteinSugra} as
\begin{eqnarray}
S_{\rm SG}^{\left(2\right)}  =  -\int d^{6}z\mathcal{W}^{\alpha\beta\gamma}\frac{1}{\square}\mathcal{W}_{\alpha\beta\gamma}-3\int d^{8}z\bar{\mathcal{R}}\frac{1}{\square}\mathcal{R}
  =  -\frac{1}{2}\int d^{8}z\,\mathcal{G}^{\alpha\dot{\alpha}}\frac{1}{\square}\mathcal{G}_{\alpha\dot{\alpha}}-\int d^{8}z\bar{\mathcal{R}}\frac{1}{\square}\mathcal{R} \, , 
  \label{eq:FieldStrenghtLinearEinsteinSugra}
\end{eqnarray}
where the following linearized field strengths (note the different math fonts for the curvatures used for linearized quantities),  
\begin{eqnarray}
&& \mathcal{W}_{\alpha\beta\gamma} = \frac{i}{8}\bar{D}^{2}\partial_{(\alpha}\,^{\dot{\beta}}D_{\beta}H_{\gamma)\dot{\beta}}\,,\quad 
 \mathcal{R}= -\frac{i}{12}\overline{D}^{2}\partial^{\beta\dot{\beta}}H_{\beta\dot{\beta}}-\frac{1}{4}\overline{D}^{2}\bar{\sigma}\,, \nonumber \\
&& \mathcal{G}_{\alpha\dot{\alpha}}=i\partial_{\alpha\dot{\alpha}}\left(\bar{\sigma}-\sigma\right)+ \frac{1}{48}\square^{-1}\left(\partial^{\beta}\phantom{}_{\dot{\alpha}}D^{\gamma}\overline{D}^{2}D_{\left(\gamma\right.}\partial_{\alpha}\phantom{}^{\dot{\beta}}H_{\left.\beta\right)\dot{\beta}}-\partial_{\alpha\dot{\alpha}}\left\{ D^{2},\overline{D}^{2}\right\} \partial^{\beta\dot{\beta}}H_{\beta\dot{\beta}}\right)  ,
\label{RGWquad}
\end{eqnarray}
are gauge invariant under the transformation (\ref{eq:LinearGaugeTransformH}) and satisfy the following Bianchi identities, 
\begin{eqnarray}
\bar{D}_{\dot{\alpha}}\mathcal{W}_{\alpha\beta\gamma}=0,\quad \bar{D}_{\dot{\alpha}}\mathcal{R}=0\,, \quad 
D^{\gamma} \mathcal{W}_{\alpha\beta\gamma}=\partial_{\left(\alpha\right.}\phantom{}^{\dot\gamma} \mathcal{G_{\left.\beta\right)\dot{\gamma}}} \,  ,       
\quad\bar{D}^{\dot{\alpha}}\mathcal{G_{\alpha\dot{\alpha}}}=D_{\alpha}\mathcal{R}\, .
\label{Bianchi}
\end{eqnarray}
The nonlocal gauge invariant generalization of \eqref{eq:FieldStrenghtLinearEinsteinSugra} is thus easily obtained introducing form factors in between the curvatures, 
\begin{eqnarray}
&& S_{\rm NL}^{\left(2\right)} =  -\int d^{6}z\mathcal{W}^{\alpha\beta\gamma}\frac{1}{\square}h_{W}\left(\square\right)\mathcal{W}_{\alpha\beta\gamma}-3\int d^{8}z\bar{\mathcal{R}}\frac{1}{\square}h_{R}\left(\square\right)\mathcal{R} \nonumber \\
 && \hspace{0.7cm} 
 =  -\frac{1}{2}\int d^{8}z\,\mathcal{G}^{\alpha\dot{\alpha}}\frac{1}{\square}h_{W}\left(\square\right)\mathcal{G}_{\alpha\dot{\alpha}}+\int d^{8}z\bar{\mathcal{R}}\frac{1}{\square}\left(2h_{W}\left(\square\right)-3h_{R}\left(\square\right)\right)\mathcal{R}\,,
 \label{eq:FieldStrenghtLinearNLSugra}
\end{eqnarray}
where $h_{W}\left(\square\right)$ and $h_{R}\left(\square\right)$ are form factors such that the spectrum is ghost-free and contains the massless superspin $3/2$ multiplet. Notice that such a spectrum is determined by only two form factors as a consequence of the linearized version of the supersymmetric Gauss-Bonnet 
theorem (\ref{SGB}), i.e\footnote{Given $(\sigma_{m})_{\alpha \dot{\alpha}}$,  
$(\tilde{\sigma}_{m})^{\dot{\alpha} \alpha} \equiv \epsilon^{\dot{\alpha} \dot{\beta}} \epsilon^{{\alpha} {\beta}} (\sigma_{m})_{\beta \dot{\beta} }$, and ${\rm Tr}( \sigma_a \tilde{\sigma}_b ) = -2 \eta_{ab}$
we find: ${\mathcal G}_a {\mathcal G}^a = {\mathcal G}^a {\mathcal G}^b \eta_{ab}
 = - \frac{1}{2} {\mathcal G}_{\alpha \dot{\alpha}} {\mathcal G}^{\dot{\alpha} \alpha}$.
 }
\begin{equation}
\frac{1}{2}\int d^{8}z\,\mathcal{G}^{\alpha\dot{\alpha}}\mathcal{G}_{\alpha\dot{\alpha}}  =  \int d^{6}z\mathcal{W}^{\alpha\beta\gamma}\mathcal{W}_{\alpha\beta\gamma}+2\int d^{8}z\bar{\mathcal{R}}\mathcal{R}\, .
\label{GB1}
\end{equation}
The action (\ref{eq:FieldStrenghtLinearNLSugra}) is the linearized expansion of the theory (\ref{SNL})
when the following identifications are plugged in (see  next subsection about the explicit linearization of (\ref{SNL})),
\be
h_{W}\left(\square\right)  = 1-\kappa^2\square \, 
\left(\gamma_W\left(\square\right) +2\gamma_G\left(\square\right) \right)\, , \quad 
 h_{R}\left(\square\right)  =  1-\frac13 \kappa^2\square\, 
\left(\gamma_{\mathcal R} \left(\square\right)+4\gamma_G \left(\square\right)  \right) ,  \quad %{\color{red} ??????}
\label{entire0}
\ee
where $\gamma_W\left(\square\right)$ and $\gamma_{\mathcal R} \left(\square\right)$ are chosen to be entire functions of their argument. We can now replace (\ref{entire0}) in the linearized action (\ref{eq:FieldStrenghtLinearNLSugra}),
\be
 \hspace{-1.3cm}
 S_{\rm NL}^{\left(2\right)}& = & -\int d^{6}z\mathcal{W}^{\alpha\beta\gamma}\frac{1}{\square}h_{W}\left(\square\right)\mathcal{W}_{\alpha\beta\gamma}-3\int d^{8}z\bar{\mathcal{R}}\frac{1}{\square}h_{R}\left(\square\right)\mathcal{R} \nonumber \\
 %&& \hspace{-1.3cm} 
 %=  -\frac{1}{2}\int d^{8}z\,\mathcal{G}^{\alpha\dot{\alpha}}\frac{1+\square \, 
%\gamma_W\left(\square\right)}{\square} 
%\mathcal{G}_{\alpha\dot{\alpha}}+\int d^{8}z\bar{\mathcal{R}}\frac{1}{\square}
%\left(2 +\square \, 
%2 \gamma_W\left(\square\right)
%-3
%-3  \square\, 
%\gamma_{\mathcal R} \left(\square\right) 
%h_{R}\left(\square\right)
%\right)
%\mathcal{R}\,, \nonumber \\
%&& \hspace{-1.3cm}
%=  -\frac{1}{2}\int d^{8}z\,\mathcal{G}^{\alpha\dot{\alpha}}\frac{1}{\square} 
%\mathcal{G}_{\alpha\dot{\alpha}}
%
%- \int d^{8}z\bar{\mathcal{R}}\frac{1}{\square}
%\mathcal{R}
%
% -\frac{1}{2}\int d^{8}z\,\mathcal{G}^{\alpha\dot{\alpha}} 
%\gamma_W \left(\square\right)
%\mathcal{G}_{\alpha\dot{\alpha}}+\int d^{8}z\bar{\mathcal{R}}
%\left(
%2 \gamma_W\left(\square\right)
%-3  
%\gamma_{\mathcal R} \left( \square \right) 
%h_{R}\left(\square\right)
%\right)
%\mathcal{R} \nonumber \\
%&& \hspace{-1.3cm}
%=  -\frac{1}{2}\int d^{8}z\,\mathcal{G}^{\alpha\dot{\alpha}}\frac{1}{\square} 
%\mathcal{G}_{\alpha\dot{\alpha}}
%
%- \int d^{8}z\bar{\mathcal{R}}\frac{1}{\square}
%\mathcal{R}
%
 %-\frac{1}{2}\int d^{8}z\,\mathcal{G}^{\alpha\dot{\alpha}} 
%\gamma_W \left(\square\right)
%\mathcal{G}_{\alpha\dot{\alpha}}
%-3 \int d^{8}z\bar{\mathcal{R}}
%\gamma_{\mathcal R} \left( \square \right) 
%h_{R}\left(\square\right)
%\mathcal{R}
%-  \int d^{6}z\mathcal{W}^{\alpha\beta\gamma}  
%\gamma_W \left(\square\right)
%\mathcal{W}_{\alpha\beta\gamma} \nonumber \\
 \hspace{-1.3cm}
&=&  
-3\int d^{8}z\bar{\mathcal{R}}\frac{1}{\square}\mathcal{R}
 -\int d^{6}z\mathcal{W}^{\alpha\beta\gamma}\frac{1}{\square}\mathcal{W}_{\alpha\beta\gamma} \\
&& +\kappa^2 \int d^{8}z\bar{\mathcal{R}}
\left(\gamma_{\mathcal R} \left(\square\right)+4\gamma_G \left(\square\right)  \right)
%h_{R}\left(\square\right)
\mathcal{R}
+\kappa^2  \int d^{6}z\mathcal{W}^{\alpha\beta\gamma}  
\left(\gamma_W\left(\square\right) +2\gamma_G\left(\square\right) \right)
\mathcal{W}_{\alpha\beta\gamma}  \, ,
\ee
where we used the Gauss-Bonnet identity (\ref{GB1}) and we missed interaction vertices.  

%$h_{G}\left(\square\right)$. 

\subsection{Unitarity}
The dynamical system \eqref{eq:FieldStrenghtLinearEinsteinSugra} is characterized by the following equations of motion, 
\begin{eqnarray}
%&&\hspace{-1cm}
 \frac{\delta S_{\rm SG}^{(2)}}{\delta H^{{\alpha\dot{\alpha}}}} =  \mathcal{G_{\alpha\dot{\alpha}}}=0\label{eq:LinearEoM} \, , \quad %\nonumber \\
%&&\hspace{-1cm}
 \frac{\delta S_{SG}^{(2)}}{\delta\sigma}  =  -3\mathcal{R}=0 \, ,
 \label{eq:EinsteinSugraLinearEoM}
\end{eqnarray}
%We see that only 
while 
the (anti-)chiral field strengths $\overline{\mathcal{W}}_{\dot{\alpha}\dot{\beta}\dot{\gamma}}$ and $\mathcal{W}_{\alpha\beta\gamma}$ do not vanish on shell, but, due to the identities \eqref{Bianchi}, they satisfy the following equations, 
\begin{equation}
D^{\alpha}\mathcal{W}_{\alpha\beta\gamma}=\bar{D}^{\dot\alpha}\overline{\mathcal{W}}_{\dot{\alpha}\dot{\beta}\dot{\gamma}}=0\, 
\end{equation}
that define the massless on-shell super-fields and the corresponding super-helicities (SH), namely 
\begin{equation}
{\rm{SH}} \left( \mathcal{W}_{\alpha\beta\gamma}\right)=\frac32, \qquad 
{\rm{SH}}  \left( \overline{\mathcal{W}}_{\dot{\alpha}\dot{\beta}\dot{\gamma}} \right)=-2 \, .
\end{equation}
Therefore, the Einstein supergravity theory, at the linearized level, describes two massless 
super-Poincar\'e states of super-helicities $-2$ and $3/2$.

%\section{Non local supergravity}
The equations of motion corresponding to the nonlocal linearized supergravity  \eqref{eq:FieldStrenghtLinearNLSugra} are a straightforward generalization of \eqref{eq:EinsteinSugraLinearEoM}
\begin{eqnarray}
&& %\hspace{-1.4cm}
\frac{\delta S_{\rm NL}^{\left(2\right)}}{\delta H^{\alpha\dot{\alpha}}\left(z\right)}  =  
\left[ \frac{h_{W}\left(\square\right)}{16\square}\left(\frac{2}{3}\left\{ D^{2},\overline{D}^{2}\right\} -2D^{\gamma}\bar{D}^{2}D_{\gamma}\right) \right.
%\mathcal{G}_{\alpha\dot{\alpha}}\left(z\right) 
 \nonumber \\
&& \hspace{1.8cm} \left. 
+\frac{1}{16\square}\left(h_{R}\left(\square\right)-\frac23 h_{W}\left(\square\right)\right)\left\{ D^{2},\overline{D}^{2}\right\} 
\right] 
\mathcal{G}_{\alpha\dot{\alpha}}\left(z\right)\nonumber\\
&& \hspace{1.8cm} =\left[-\frac{h_{W}\left(\square\right)}{8\square}D^{\gamma}\bar{D}^{2}D_{\gamma}+\frac{h_{R}\left(\square\right)}{16\square}\left\{ D^{2},\overline{D}^{2}\right\} \right]
\mathcal{G}_{\alpha\dot{\alpha}}\left(z\right)=0\, , \label{eq:NLSugraLinearEoMG} \\
&&  %\hspace{-1.4cm}
\frac{\delta S_{NL}^{\left(2\right)}}{\delta{\sigma}\left(z\right)} =  -3h_{R}\left(\square\right){\mathcal{R}}\left(z\right)=0 \, .
\label{eq:NLSugraLinearEoMR}
\end{eqnarray}
We can infer from the local supergravity case what is fundamental requirement to get the on-shell Einstein supergravity spectrum, namely the field strengths ${\mathcal{R}}$ and $\mathcal{G}_{\alpha\dot{\alpha}}$ must be analytical functions in the momentum space without any extra poles corresponding to physical propagating degrees if freedom.  This is tantamount to requiring that the differential operators acting on  ${\mathcal{R}}$ and $\mathcal{G}_{\alpha\dot{\alpha}}$ have a well defined inverse for any value of $\square=-k^2$ in momentum space. Such inverse can be easily constructed in the basis of projectors
\begin{equation}
\mathscr{P}_{(0)}=-\frac{1}{8}\frac{D^{\gamma}\bar{D}^{2}D_{\gamma}}{\square}\,, 
\quad 
\mathscr{P}_{(+)}=\frac{1}{16}\frac{\overline{D}^{2}D^{2}}{\square}\,,
\quad 
\mathscr{P}_{(-)}=\frac{1}{16}\frac{{D}^{2}\overline{D}^{2}}{\square}\,,
\end{equation}
satisfying the identities
\begin{eqnarray}
&&\mathscr{P}_{(i)}\mathscr{P}_{(j)}=\delta_{ij}\mathscr{P}_{(i)}\,,\label{eq:Proj1}\\
&&\mathscr{P}_{(0)}+\mathscr{P}_{(+)}+\mathscr{P}_{(-)}=\mathbb{I}\, .
\label{eq:Proj2}
\end{eqnarray}
In fact, the equations \eqref{eq:NLSugraLinearEoMG} and \eqref{eq:NLSugraLinearEoMR}
can be rewritten as
\begin{eqnarray}
&& \left[{h_{W}\left(\square\right)}\mathscr{P}_{(0)}+{h_{R}\left(\square\right)}\left(\mathscr{P}_{(+)}+\mathscr{P}_{(-)}\right) \right]
\mathcal{G}_{\alpha\dot{\alpha}}\left(z\right)=0\, , \nonumber \\
&&-3h_{R}\left(\square\right){\mathcal{R}}\left(z\right)=0 \, , 
\end{eqnarray}
and the inverses can be constructed in the form 
\be
\alpha_{G,R}\mathscr{P}_{(0)}+\beta_{G,R}\left(\mathscr{P}_{(+)}+\mathscr{P}_{(-)}\right).
\ee
 Using the identities \eqref{eq:Proj1} and \eqref{eq:Proj2} it is straightforward to find the solutions
\begin{eqnarray}
&&\alpha_{G}={h_{W}^{-1}\left(\square\right)}\,, \quad \beta_{G}={h_{R}^{-1}\left(\square\right)}\,, \nonumber \\
&& \alpha_R=\beta_R=-\frac13 h_{R}^{-1}\left(\square\right).
\end{eqnarray}
which are well defined for any  $\square=-k^2$ only if $h_{W}\left(\square\right)$ and $h_{R}\left(\square\right)$ have no poles.

Therefore, if we want to get the graviton and the gravitino  kinetic terms with the standard normalizations, we should assume the following form factors, 
\begin{equation}
 \quad h_W(\Box) =  e^{{\rm H}_2(\Box)} \, , \quad h_R(\Box) =e^{{\rm H}_0(\Box)} \, ,
  \end{equation}
  and still we have the freedom to choose the two entire functions ${\rm H}_2$ and ${\rm H}_0$.  
In terms of the form factors in (\ref{SNL}) we have 
\be
\gamma_{W}\left(\square\right) = -\frac{e^{{\rm H}_2(\square)}-1}{\kappa^2\square}-2\gamma_G\left(\square\right)\,, \quad 
\gamma_{\mathcal R} \left(\square\right)=-3\frac{e^{{\rm H}_0(\square)}-1}{\kappa^2\square}-4\gamma_G\left(\square\right)\,,
\label{eq:GoodFormFactors}
\ee
where ${\rm H}_0(\square)$ and ${\rm H}_2(\square)$ are entire functions on the complex plane that can be taken as described in \cite{Tombo,modesto,modestoLeslaw}.
The easiest choice is
\begin{equation}
 \quad h_W(\Box) = h_R(\Box) = e^{\rm H(\Box)} \, , 
\end{equation}
or equivalently,
\be
\gamma_{W}\left(\square\right) = -\frac{e^{{\rm H}(\square)}-1}{\kappa^2\square}-2\gamma_G\left(\square\right)\,, \quad 
\gamma_{\mathcal R} \left(\square\right)=-3\frac{e^{{\rm H}(\square)}-1}{\kappa^2\square}-4\gamma_G\left(\square\right)\,,
\label{eq:GoodFormFactorsH}
\ee
where again ${\rm H}(\square)$ is an entire function on the complex plane that can be taken as described in \cite{Tombo, modesto,modestoLeslaw} (one explicit example will be given later in section five (\ref{FormFactor}).)
The corresponding equations of motion are in this case
\begin{eqnarray}
&&e^{{\rm H}(\square)} \mathcal{G}_{\alpha\dot{\alpha}}\left(z\right)=e^{{\rm H}(\square)}{\mathcal{R}}\left(z\right)=0\, ,
\label{eq:OneEntireNLSugraLinearEoM}
\end{eqnarray}
which leave the only on-shell dynamical fields in the supermultiplets $\overline{\mathcal{W}}_{\dot{\alpha}\dot{\beta}\dot{\gamma}}$ and $\mathcal{W}_{\alpha\beta\gamma}$ in complete analogy with the Einstein supergravity case.

The same unitarity %{\color{red}on-shell} 
analysis can be done rewriting the action (\ref{eq:ProjLinearEinsteinSugra})
%expanding the action {\color{red}WHAT ACTION?} 
in the components
%that can be obtained in the Wess-Zumino gauge defined by equations 
%that are 
defined in \eqref{eq:WessZumino} and \eqref{eq:WessZuminoComp} in the Wess-Zumino gauge.
The following relationships have to be implemented 
in order to obtain the Lagrangian quadratic in the component fields, 
\begin{eqnarray*}
&& -\frac{1}{4} \bar{D}^{2}D_{\beta}H_{\alpha\dot{\alpha}} \Big| =    i\Psi_{\alpha\dot{\alpha},\beta}=i\left(\sigma^{a}\right)_{\alpha\dot{\alpha}}\Psi_{a\beta}\, , \quad %\nonumber \\&&
 \frac{1}{32}\left|\left\{ D^{2},\bar{D}^{2}\right\} H_{\alpha\dot{\alpha}}\right|  =  A_{\alpha\dot{\alpha}}=\left(\sigma^{a}\right)_{\alpha\dot{\alpha}}A_{a}\, , \nonumber \\
&& \frac{1}{2} \left[D_{\beta},\bar{D}_{\dot{\beta}}\right]H_{\alpha\dot{\alpha}}\Big| 
  =  h_{\beta\dot{\beta},\alpha\dot{\alpha}}=\left(\sigma^{b}\right)_{\beta\dot{\beta}}\left(\sigma^{a}\right)_{\alpha\dot{\alpha}}h_{ba}\, , \nonumber \\
&&
\bar{\sigma}\Big| 
  =  \frac{1}{6}h_{\alpha\dot{\alpha}}\phantom{}^{\alpha\dot{\alpha}}=-\frac{1}{3}h_{a}^{a}\, ,\quad 
 \bar{D}_{\dot{\alpha}}\bar{\sigma}\Big| 
  =  \frac{2}{3}i\Psi_{\alpha\dot{\alpha},}\phantom{}^{\alpha}=\frac{2}{3}i\left(\sigma^{a}\right)_{\alpha\dot{\alpha}}\Psi_{a}\phantom{}^{\alpha}\, \, , \quad 
  -\frac{1}{4}\bar{D}^{2}\bar{\sigma}\Big| \, 
  =  \frac{1}{3}\bar{B}\, .
\end{eqnarray*}
With the definitions above the expression of the linearized Einstein supergravity Lagrangian 
$\mathcal{L}_{\rm SG}^{\left(2\right)}$ in components, which corresponds to the superspace linearized action \eqref{eq:LinearEinsteinSugra}, reads 
\begin{eqnarray}
&& \mathcal{L}_{\rm SG}^{\left(2\right)}  =  -\frac{1}{2}\left(\partial^{c}h^{ab}\right)\partial_{c}h_{ab}+\frac{1}{2}\left(\partial^{c}h_{a}^{a}\right)\partial_{c}h_{b}^{b}+\left(\partial_{b}h^{ab}\right)\left(\partial^{c}h_{ac}-\partial_{a}h_{c}^{c}\right)
\nonumber \\
&& \hspace{1.1cm}
  -\frac{1}{3}\bar{B}B+\frac{4}{3}A^{m}A_{m}+\frac{\epsilon^{abcd}}{2} \overline{\Psi}_{a} 
  \tilde{\sigma}_{b} 
  \Psi_{cd}\, , \label{eq:LinearEinsteinSugraLagrangian}
\end{eqnarray}
where $\Psi_{cd\alpha}=\partial_{c}\Psi_{d\alpha}-\partial_{d}\Psi_{c\alpha}$.
  The first three operators in \eqref{eq:LinearEinsteinSugraLagrangian} come from the linearized Einstein-Hilbert gravity action when only terms quadratic in the fluctuation field $h_{ab}$ ($g_{ab}={\eta}_{ab}-h_{ab}/2$) are kept. It is well known that such an action describes a propagating massless spin-$2$ particle. 
The fields $B$ and $A_m$ have trivial dynamics due to their equation of motions, namely 
\begin{equation}
B=A_m=0\, ,
\end{equation}
and this is the reason why they are usually called supergravity auxiliary fields. 
Finally, the last term in (\ref{eq:LinearEinsteinSugraLagrangian}) is the Rarita-Schwinger action for a massless spin-$\frac32$ particle.

It is quite straightforward to derive the analogous Lagrangian $\mathcal{L}_{\rm NL}^{\left(2\right)}$ for the action \eqref{eq:FieldStrenghtLinearNLSugra}  in the case
$h_{W}\left(\square\right)=h_{R}\left(\square\right)=\exp {{\rm H} \left(\square\right)}$. The outcome is: 
\be
&& \mathcal{L}_{\rm NL}^{\left(2\right)}  =  -\frac{1}{2}\left(\partial^{c}h^{ab}\right)e^{H\left(\square\right)}\partial_{c}h_{ab}+\frac{1}{2}\left(\partial^{c}h_{a}^{a}\right)e^{H\left(\square\right)}\partial_{c}h_{b}^{b}+\left(\partial_{b}h^{ab}\right)e^{H\left(\square\right)}\left(\partial^{c}h_{ac}-\partial_{a}h_{c}^{c}\right) \nonumber \\
 && \hspace{1.1cm}
 -\frac{1}{3}\bar{B}e^{H\left(\square\right)}B+\frac{4}{3}A^{m}e^{H\left(\square\right)}A_{m}+\frac{1}{2}\epsilon^{abcd}\overline{\Psi}_{a}\tilde{\sigma}_{b}e^{H\left(\square\right)}\Psi_{cd}\, .
 \label{eq:LinearEinsteinSugraLagrangianNL}
\ee
Comparing this action with the local one (\ref{eq:LinearEinsteinSugraLagrangian}) we see that the fields $B$ and $A_m$ have the following equations of motion, 
\be
&& e^{H\left(\square\right)} B= 0 \, , \nonumber \\
&& e^{H\left(\square\right)}A_m =0\, ,
\ee
and again we have 
a trivial dynamics whether ${\rm H}(\square)$ is chosen to be an entire function. 
Therefore, the fields $B$ and $A_m$ are 
non propagating fields as in local supergravity. 
The remaining terms provide the nonlocal generalization of the massless spin2 and Rarita-Schwinger Lagrangians that, for the chosen form factors, describe the propagation of a massless spin-$2$ and a massless spin-$\frac32$ particle respectively.

\subsection{Linearizing (\ref{SNL})}
We can also get the linearized action \eqref{eq:FieldStrenghtLinearNLSugra} applying a top-down procedure, namely we can directly expand the proposed theory (\ref{SNL}).  
%%%my addition%%%
For this achievement 
%In order to do so 
we will use the prepotential parametrization for whose definition we refer to the section 5.6 of \cite{Buchbinder}. It will be convenient to work in the chiral representation that allows to express the superfield curvatures in terms of the prepotentials $H^a=\overline{H}^a$, $\varphi$ and $\overline\varphi$, with $\varphi$ being a flat chiral superfield, $\overline{D}_{\dot\alpha}\varphi=0$.  In particular,  one can compute the first order variation of geometrical
quantities constructed from the covariant derivatives 
%{$\tilde{\nabla}_{A}$} 
to the first order in the fluctuation superfields ${H}$ and $\sigma\left(\mathbf{\varphi}=e^{\sigma}\right)$,
i.e. considering the approximation
\be
e^{-2i{H}}\approx1-2i{H}^{a}\partial_{a} \, , \qquad\mathbf{\varphi}=e^{\sigma}\approx1+\sigma  .
\ee
The canonical dimensions of the fields defined in this way are $\left[{H}^{a}\right]=-1$ and $\left[\sigma\right]=0$, which means they are obtained from the ones define in the previous section by the rescalings: ${H}^{a}\rightarrow\frac{1}{\kappa}{H}^{a}$ and $\sigma\rightarrow\frac{1}{\kappa}\sigma$. One can find the linearized expressions for the superfield strengths 
\begin{eqnarray}
\tilde{W}_{ \alpha\beta\gamma}=\kappa\mathcal{W}_{\alpha\beta\gamma},\quad \tilde{G}_{\alpha\dot\alpha}=\kappa\mathcal{G}_{\alpha\dot{\alpha}}, 
\quad \tilde{R}=\kappa\mathcal{R}\,,
\end{eqnarray}
which can be written in terms of the vector superfield
$
{H_{\alpha\dot\alpha}^{\prime}} = H _{\alpha\dot\alpha}+\frac{3i}{2}\frac{\partial_{\alpha\dot\alpha}}{\square}\left(\bar{\sigma}-\sigma\right)
$,
\be 
%\hspace{-0.6cm} 
 && {\mathcal{R}}  =  -\frac{i}{12\kappa}\bar{D}^{2}\partial^{\beta\dot\beta}\Pi_{0}^L {H^{\prime}}_{\beta\dot\beta} \, , \qquad 
{\mathcal{G}}_{\alpha\dot{\alpha}}  =  
-\frac1\kappa\square (\Pi_{\frac{3}{2}}^T-\frac{2}{3}\Pi_0^L)H _{\alpha\dot\alpha}^{\prime}  \, , \nonumber \\
&&
 {\mathcal{W}}_{\alpha\beta\gamma}  =  \frac{i}{8\kappa}\bar{D}^{2}\partial_{(\alpha}\,^{\dot{\beta}}D_{\beta}\Pi_{\frac{3}{2}}^T H_{\gamma)\dot{\beta}}\,.
\ee
The quadratic action for the pure Einstein-Hilbert 
supergravity (\ref{eq:ProjLinearEinsteinSugra}) in terms of the superfield 
$H^\prime_{\alpha\dot\alpha}$ reads
\be
S_{\rm SG}^{\left(2\right)}=\frac1{\kappa^2}\int d^{8}z\, H^{\prime\,\alpha\dot{\alpha}}\square\left(-\frac{1}{3}\Pi_{0}^{L}+\frac{1}{2}\Pi_{\nicefrac{3}{2}}^{T}\right)H^\prime_{\alpha\dot{\alpha}}\, . 
  \label{EHsgra2}
\ee
In general for any higher derivative operator quadratic in the curvature superfields, like the ones introduced in (\ref{SNL}), it is sufficient to expand them to the first order 
to get an action quadratic in $H_{\alpha\dot\alpha}'$. Therefore, %It turns out that
for the form factors $\gamma_{R}$, $\gamma_{G}$,
and $\gamma_{W}$ only the zero order in $H_{\alpha\dot\alpha}'$ contributes 
to the quadratic action, and we
can substitute the curved superspace d'Alembertian with their flat
counterpart. Finally, the quadratic operators in (\ref{SNL}) at the second order in $H_{\mu}'$ reads
\be
&& %\hspace{-2cm} 
 S^{(2)}_{{\mathcal R}} =  \int d^{8}zE^{-1}{ R} \,  \gamma_{{R}}(\Delta_{{ R}}){ \bar{R}} %\right)^{(2)} %\nonumber\\&=&
=  \int d^{8}z{\tilde E}^{-1}{\tilde R} \,  \gamma_{{ R}}(\tilde \Delta_{{ R}}){\tilde {\bar{R}}} %\right)^{(2)} %\nonumber\\&=&
= \kappa^2 \int d^{8}z\bar{\mathcal{R}} \gamma_{\mathcal R} \left(\square\right) \mathcal{R}\nonumber \\
&&
%\nonumber\\&=& 
\hspace{0.7cm} 
= \frac{1}{9}\int d^{8}z H^{\prime\,\alpha\dot{\alpha}}\square^2 \gamma_{\mathcal R} \left(\square\right)\Pi_{0}^{L}H^\prime_{\alpha\dot{\alpha}} \, ,  \\
&& 
S^{(2)}_{\rm G} =  \int d^{8}z E^{-1}G^{\alpha\dot\alpha} \, \gamma_{G}(\Delta_{G})G_{\alpha\dot\alpha}  %\right)^{(2)} %\nonumber\\&=& 
= \int d^{8}z {\tilde E}^{-1}{\tilde G}^{\alpha\dot\alpha} \, \gamma_{G}(\tilde \Delta_{G}){\tilde G}_{\alpha\dot\alpha}  %\right)^{(2)} %\nonumber\\
%&=& 
=\kappa^2 \int d^{8}z\,\mathcal{G}^{\alpha\dot{\alpha}}\gamma_G\left(\square\right)\mathcal{G}_{\alpha\dot{\alpha}}\nonumber\\
&& \hspace{0.7cm}
= \int d^{8}z H^{\prime \, \alpha\dot\alpha}\, \square^{2} 
\, 
\gamma_{G}(\square) \left(-\Pi^T_{\frac{3}{2}} + \frac{4}{9} \Pi^L_{0}\right) H _{\alpha\dot\alpha}^{\prime}\, , \\
&& S^{(2)}_{\rm W} =  \int d^{6}z \frac{E^{-1}}{{ R}} W^{\alpha\beta\gamma} \, \gamma_{W}(\Delta_{W}) \, W_{\alpha\beta\gamma}% \right)^{(2)}\nonumber\\
= \int d^{6}z {\hat \varphi}^3 {\tilde W}^{\alpha\beta\gamma} \, \gamma_{W}(\tilde \Delta_{W}) \, {\tilde W}_{\alpha\beta\gamma}% \right)^{(2)}
\nonumber\\
&& \hspace{0.7cm} 
=- \frac12 \int d^{8}z H^{\prime \, \alpha\dot\alpha} \, \square^{2} \, \gamma_{W}(\square) \Pi^T_{\frac{3}{2}}H _{\alpha\dot\alpha}^{\prime} \, .
\ee

Collecting together the quadratic expansions for the Einstein-Hilbert and the higher derivative operators 
we finally get the quadratic expansion for the nonlocal supergravity (\ref{SNL}) in superspace, 
%The quadratic action for a general higher derivative supergravity
%of the kind discussed here is 
\be
S_{\rm NL}^{(2)} = S_{\rm SG}^{(2)} + S^{(2)}_{\mathcal R} + S^{(2)}_{\rm G} + S^{(2)}_{\rm W} = 
% S_{\rm HD}^{(2)} = 
\int d^{8}zH^{\prime \, \alpha\dot\alpha}\square\left[h_{\frac{3}{2}}(\square)\Pi^T_{\frac{3}{2}}+h_{0}(\square)\Pi^L_{0}\right]H _{\alpha\dot\alpha}^{\prime} \, , 
\label{SNL2}
\ee
where we introduced the following definitions, 
\be
&& h_{\frac{3}{2}}(\square)  =  \frac{1}{2\kappa^{2}}-\square\left(\gamma_{G}(\square)+\frac12\gamma_{W}(\square)\right) , \nonumber  \\
&& h_{0}(\square)  =  -\frac{1}{3\kappa^{2}}+\frac{1}{9}\square\left(4\gamma_{G}(\square)+\gamma_{{ R}}(\square)\right)  .
\label{entire}
\ee
In order to %have the same spectrum of the $N=1$ supergravity and 
retain unitarity, the above functions (\ref{entire})
must be entire functions with no zeros in the all complex plane.

Comparing (\ref{SNL2}) with (\ref{EHsgra2}) 
the most general choice for the form factors compatible with unitarity is the following,
\be
&&  \frac{1}{2\kappa^{2}} V_2(  \Box) \equiv  h_{\frac{3}{2}}(\square) 
 =   \frac{1}{2\kappa^{2}}-\square\left(\gamma_{G}(\square)+\frac12\gamma_{W}(\square)\right)  , \\
&&  -\frac{1}{3\kappa^{2}} V_0(  \Box) \equiv h_{0}(\square)  
=  -\frac{1}{3\kappa^{2}}+\frac{1}{9}\square\left(4\gamma_{G}(\square)+\gamma_{{ R}}(\square)\right)  .
\label{entireV}
\ee
We must assume $V_2( 0) = V_0(  0 ) =1$ to have the same residue of local supergravity in $-\Box = k^2 =0$.  
Solving for two out of the three form factors $\gamma_{\mathcal{R}}$, $\gamma_{{G}}$, and $\gamma_{W}$
we can write the action (\ref{SNL}) in terms of $V_2( \Box)$, $V_0( \Box)$ and the remaining form factor. The solution is in agreement with (\ref{eq:GoodFormFactorsH}). 
In particular assuming the minimal choice $V_0(\Box) = V_2(\Box) \equiv e^{{\rm H}(\Box)}$ the quadratic action reads
\be
S_{\rm NL}^{(2)} = 
\frac{1}{\kappa^{2}}\int d^{8}z H^{\prime \, \alpha\dot\alpha} \, \square \, e^{{\rm H}(\Box)} \left(\frac12\Pi^T_{\frac{3}{2}}-\frac{1}{3}\Pi^L_{0} \right) H _{\alpha\dot\alpha}^{\prime}  \, .
\label{SNL3}
\ee

The kinetic operator of the nonlocal supergravity is the same of the local one, but it is multiplied by a form factor,
which makes the
theory more convergent in the ultraviolet regime without changing the spectrum or introducing poltergeist states.
In short, the propagators for the graviton and the gravitino have both the following simplified structure,  
\be
{\mathcal O}^{-1}_{\rm NL} =  e^{- {\rm H}(\Box)}  {\mathcal O}^{-1}_{\rm SG} =  \frac{e^{ -{\rm H}(\Box)} }{\Box} \times {\mbox{tensor structure} }\, .
\ee
%where TS means tensor structure. 
%
%Expanding the action in components, %matter content of the theory is the same of the 
%Let us start with the simple $N=1$ supergravity in four spacetime dimensions.
%the supersymmetric multiplet consists on the spin-$2$ graviton $h_{\mu\nu}$, the spin-$3/2$ gravitino $%\psi_{\mu}$, and three auxiliary fields $S, P,A_\mu$ with mass dimension: $[h_{\mu \nu}] = 0$,  
%$[\psi_\mu] = 3/2$, $[S] = [P] = [A_{\mu}] = 2$. 
The propagators for the component fields can be read from the 
components of $H_{\mu}'$ and using the action (\ref{SNL3}) and/or (\ref{eq:LinearEinsteinSugraLagrangianNL})
\be
&& \langle h \,h \rangle \propto %= 8 
\frac{e^{-H(\Box)}}{ \Box}  \left[ P^2 - \frac{P^{0,s}}{2} \right]  \, , 
 \, \quad
 \langle \psi \, \psi \rangle \propto %= - 2
  \frac{e^{-H(\Box)}}{ \not\!\partial}   
\left[ P^{3/2} - 2 P^{1/2} \right] \, , \nonumber  \\
&& 
\langle A \, A \rangle  \propto %= 3 
e^{-H(\Box)} 
\left[ P^1 - %era piu ma non puo essere giusto
 P^{0} \right]  %\gamma^{-1} 
\, ,  \label{Af} \, \quad 
\langle B \, \bar{B} \rangle %= \langle P \, P \rangle \propto %= - 3 
\propto e^{-H(\Box)} % \, \gamma^{-1} 
\label{SPf} \,.
\ee
In (\ref{SPf}) all the indices have been omitted and  
the projectors, which satisfy orthonormality, decomposition of the unity and completeness \cite{VN}, are: 
\be
&& P^{3/2}_{\mu \nu} = \theta_{\mu \nu} -  \frac{1}{3} \hat{\gamma}_{\mu} \hat{\gamma}_{\nu} \,  , \quad 
\hat{\gamma}_{\mu} = {\gamma}_{\mu} - \omega_{\mu} \, , 
 \quad
 (P^{1/2}_{11} )_{\mu \nu} = \frac{1}{3} \hat{\gamma}_{\mu} \hat{\gamma}_{\nu} \,  , \quad 
 (P^{1/2}_{12} )_{\mu \nu} = \frac{1}{\sqrt{3}} \hat{\gamma}_{\mu}\omega_{\nu} \,  ,
 \nonumber \\
 &&
 (P^{1/2}_{21} )_{\mu \nu} = \frac{1}{\sqrt{3}} \omega_{\mu} \hat{\gamma}_{\nu} \,  ,
 \quad 
 (P^{1/2}_{22} )_{\mu \nu} = \omega_{\mu}\omega_{\nu} \,  , \quad 
  \theta_{\mu \nu} = \delta_{\mu \nu} - \omega_{\mu} \omega_{\nu} \,  , \quad 
  \omega_{\mu} = \frac{\partial_{\mu} \not\!\partial}{\Box} \, .
\ee
$P^{2}, P^{0,s}, P^{0,ts}$ %, P^{(0-\omega s)} $ 
are the spin-$2$ projectors defined in \cite{VN} and $P^1$ and $P^0$ the vector field projectors.

Once more, the graviton and the gravitino have only one pole in $ -\Box =0$, while the auxiliary fields 
do not propagate at all because their two-point functions have no poles just as in the local theory.

We give here the last consistency check based on the Gauss-Bonnet identity.  
Inverting the relations (\ref{entireV}) for the form factors defined in (\ref{SNL}) we find the following expressions, 
\be
&& 
\gamma_W = -\frac{1}{\kappa^2} \frac{e^{\rm H} -1}{\Box} - 2\gamma_G \, ,  \nonumber \\ %\quad
&& 
\gamma_{ R} = - \frac{3}{\kappa^2} \frac{e^{\rm H} -1}{\Box} -4  \gamma_G \, ,
\ee
which coincide with the expressions \eqref{eq:GoodFormFactors} for 
${\rm H}_2 = {\rm H}_0 \equiv {\rm H}$.
When the above form factors are plugged in the action (\ref{SNL}) we find the following final expression,
\be
%\hspace{-0.0cm}
&& %\hspace{-0.3cm} 
S_{\rm NL} \! = \! \frac{1}{\kappa^2} \int \! d^8 z E^{-1} \left[ - 3   
+  {R} \left(  -3 \frac{e^{{\rm H}(\Delta_{{R}})} - 1}{\Delta_{{R}}}  - 4 \kappa^2 \gamma_G \right) \, \bar{{R}}+ \kappa^2 {G}^{\alpha\dot\alpha} \,  \gamma_{G}  \, { G}_{\alpha\dot\alpha} \right] \label{SNLF} \\
&& %\hspace{-0.3cm}
+ \frac{1}{\kappa^2}
\int \! d^8 z  \frac{E^{-1}}{R} 
W_{\alpha \beta \gamma} \, \left( - \frac{e^{\rm H\left(\Delta_W \right)} -1}{\Delta_W} - 2\kappa^2\gamma_G \right) \, W^{\alpha \beta \gamma} .
\!\!\! \nonumber  \\
&&  %\hspace{-1cm} \hspace{0.7cm}
 =
\frac{1}{\kappa^2} \int \! d^8 z E^{-1} \left[ - 3   
- 3  {R} \left(   \frac{e^{{\rm H}(\Delta_{{R}})} - 1}{\Delta_{{R}}}  \right) \, \bar{{R}} \right]  %\nonumber \\
%&&  \hspace{0.7cm}
+ \frac{1}{\kappa^2}
\int \! d^8 z  \frac{E^{-1}}{R} 
W_{\alpha \beta \gamma} \, \left( - \frac{e^{\rm H\left(\Delta_W \right)} -1}{\Delta_W} \right) \, W^{\alpha \beta \gamma} 
\nonumber \\
&& %\hspace{-0.3cm} 
+ 2\underbrace{ \int \! d^8 z E^{-1} \left[
-2 \bar{R}   \gamma_G\left(\Delta_R\right) {R} +\frac12  { G}^{\alpha\dot\alpha} \,  \gamma_{G}\left(\Delta_G\right)   \, {G}_{\alpha\dot\alpha} \right] +
2\int \! d^8 z  \frac{E^{-1}}{R} 
W_{\alpha \beta \gamma} \, 
\gamma_G \left(\Delta_W\right)  \, 
W^{\alpha \beta \gamma}}_{{\rm GB}_{\rm nl}}  \, , 
\nonumber 
\ee
where ${\rm GB}_{\rm nl}$ is the ``delocalized" Gauss-Bonnet that only contribute to vertices
without touching unitarity. The replacement of the form factors in (\ref{SNLF}) is a further check 
of the computation, because only two out of three form factors can contribute to the propagator
as explicitly shown by the reconstruction of the Gauss-Bonnet operator.

\section{Nonlocal supergravities with modified spectrum}
In this section we provide the supersymmetric extension of the weakly nonlocal Starobinsky theory 
proposed in \cite{star} and we construct two other nonlocal supergravity theories
based on the Bosonic massive gravity derived in \cite{TsujikawaModestoIR}. 
The second one (see section (\ref{Pauli})) is here proposed for the first time also at the Bosonic level, which 
consists on exactly the five degrees of freedom of the Pauli-Firtz massive gravity.

\subsection{Nonlocal supergravity completion of the Starobinsky theory}
In our nonlocal framework unitarity is achieved by requiring that only the graviton and gravitino are propagating degrees of freedom. So one removes not only the ghost-like particle which is the most serious blight of Stelle's theory, but also 
the scalar field which is crucial for the Starobinsky's model of  inflation \cite{star}. 
In the non-supersymmetric case  a purely gravitational super-renormalizable completion of the Starobinsky theory has been proposed in \cite{Briscese:2013lna}. In this section we will 
make a choice of the form factors (\ref{entire}) in order to have the supersymmetric analogue of the result in \cite{Briscese:2013lna}.
The supersymmetric completion of the Starobinsky theory is obtained by solving (\ref{entireV})
for $\gamma_W$ and $\gamma_{R}$ (we take $\gamma_G =0$ for the sake of simplicity),
%but for different fall off entire function $V_0$ and $V_2$, 
%
\be
\gamma_W = - \frac{ 1}{\kappa^2} \frac{e^{\rm H_2(\Delta_W)} -1}{\Delta_W} -2  \gamma_G \, , \quad
\gamma_{ R} = - \frac{3}{\kappa^2} \frac{e^{\rm H_0(\Delta_R)} -1}{\Delta_R} -4  \gamma_G \, . 
\label{gammaWR}
\ee
The most general spectrum compatible with unitarity is achieved by the following replacement,
\be
e^{{\rm H}_0(\Delta_R)} \,\,\,  \rightarrow \, \,\, e^{{\rm H_0(\Delta_R)}} \left(  1 - \frac{\Delta_R}{ m^2} \right),
\ee
where now it is the product $\Delta_R \, \exp H_0(\Delta_R) $ to have the same ultraviolet 
fall-off of the entire function $\exp H_2$.
Finally, the action reads
%%%
\be \hspace{-0.6cm}
S_{\rm S} = 
 \frac{- 3}{\kappa^2} \int \! d^8 z  E^{-1} \left[ 1
+    {R}  \left(
\frac{e^{{\rm H}_0(\Delta_R)} \left(1 -\frac{ \Delta_R }{ m^2} \right) - 1}{\Delta_R}  \right) 
 \bar{{R}}  %\nonumber \\
%&&  \hspace{0.7cm}
+ %\frac{1}{\kappa^2}
  \frac{1}{3 {R} } %\hat{\phi}^3 
W_{\alpha \beta \gamma} \left(  
\frac{e^{{\rm H}_2(\Delta_W)} -1}{\Delta_W} \right) 
 W^{\alpha \beta \gamma} \right] %+ {\rm h.c}. 
 \! .
\ee
The introduction of the scalar degree of freedom in the Bosonic spectrum enlarge consistently the supersymmetric multiplet as well. 
%{\color{red}Indeed the spectrum now consist on: }
In particular the linearized theory can be read out of (\ref{SNL3}), 
\be
S_{\rm S}^{(2)} = 
\frac{1}{2 \kappa^{2}}\int d^{8}z H^{\prime \, \alpha\dot\alpha} \, \square \, \left[ 
e^{{\rm H}_2(\Box)} 
\Pi^T_{\frac{3}{2}}
-\frac{2}{3}
e^{{\rm H}_0(\Box)}  \left(1 -\frac{ \Box }{ m^2}   \right) 
\Pi^L_{0}
 \right] H _{\alpha\dot\alpha}^{\prime}  \,  .
\label{SNL3St}
\ee
Looking at the linearization of the action in terms of the super-metric $H _{\alpha\dot\alpha}^{\prime}$
we can infer the spectrum of the theory. Since the extra massive pole appears together the 
spin zero projector $\Pi^L_{0}$ the Bosonic spectrum consist on: the Starobinsky scalaron coming from 
the spin zero sector of the graviton field, one complex auxiliary scalar field $B$, and the spin zero mode of the auxiliary field $A_m$ that all now propagate. The Fermionic partners come from the
spin zero sector of the gravitino field.  
This analysis results much clearer looking at the propagators (up to irrelevant multiplicative factors)
for the component fields. %, that can be read 
%Expanding the action in components, %matter content of the theory is the same of the 
%Let us start with the simple $N=1$ supergravity in four spacetime dimensions.
%the supersymmetric multiplet consists on the spin-$2$ graviton $h_{\mu\nu}$, the spin-$3/2$ gravitino $\psi_{\mu}$, and three auxiliary fields $S, P,A_\mu$ with mass dimension: $[h_{\mu \nu}] = 0$,  
%$[\psi_\mu] = 3/2$, $[S] = [P] = [A_{\mu}] = 2$. 
The propagators %for the component fields 
can be obtained expanding the linearized action 
%components of 
%from $H_{\mu}'$ in 
(\ref{SNL3St}) in component fields and generalizing the analysis of %s in the same way of 
section (\ref{SNL}) to the case of two form factors, 
\be
&& \langle h \,h \rangle \propto %= 8 
\frac{e^{-{\rm H}_2 (\Box)}}{ \Box}  P^2 - \frac{e^{-{\rm H}_0(\Box)}}{ \Box(1-\Box/m^2)} \frac{P^{0,s}}{2}   \, , 
 \, \quad
 \langle \psi \, \psi \rangle \propto %= - 2
  \frac{e^{-{\rm H}_2 (\Box)}}{ \not\!\partial}   
P^{3/2} -  \frac{e^{-{\rm H}_0(\Box)}}{ \not\!\partial (1-\Box/m^2)} 2 P^{1/2}  \, , \nonumber  \\
&& 
\langle A \, A \rangle  \propto 
e^{-{\rm H}_2 (\Box)} P^1 
- \frac{e^{- {\rm H}_0 (\Box)}}{(1-\Box/m^2) }    P^{0}  %\gamma^{-1} 
\, ,   \, \quad 
\langle B \, \bar{B} \rangle \propto 
\frac{e^{-{\rm H}_0(\Box)}}{ \Box - m^2} 
\label{SPf2} \,.
\ee
Finally, the multiplet consist on \cite{Ferrara:1978rk}: 
(i) the usual massless spin-$(2,3/2)$ multiplet; 
(ii) one massive $(1/2, 0^+, 0^-)$ multiplet of mass $m$ given by the scalars 
$B$, $\bar{B}$ and one of the spin-$1/2$ components of the gravitino field; 
(iii) one massive $(1/2, 0^+, 0^-)$ multiplet of mass $m$ given by the spin-$0$ sector of the vector,
the Starobinsky scalaron, and the other spin-$1/2$ component of the gravitino field.

\subsection{Nonlocal Massive Supergravity}
It is straightforward to generalize the Bosonic nonlocal massive gravity \cite{Maggiore, Deser, Odintsov}, \cite{TsujikawaModestoIR} to a general covariant and supesymmetric theory.
Let us first solve (\ref{SNL3}) for $\gamma_G$ and $\gamma_{R}$ out of the three form factors,
\be
&& \gamma_G = - \frac{1}{2 \kappa^2} \frac{e^{\rm H(\Delta_G)} -1}{\Delta_G} -  \frac{ \gamma_G}{2} \, , %\quad 
\nonumber \\
&&
\gamma_{R} = -\frac{1}{\kappa^2} \frac{e^{\rm H(\Delta_R)} -1}{\Delta_R} + 2  \gamma_W \, .
\label{massiveForm}
\ee
The reason we have solved for $\gamma_G$ and $\gamma_{R}$ relies in our previous result for the 
Bosonic massive gravity. Indeed $G_\mu$ is the analog of $R_{\mu\nu}^2$. 
The minimal diffeomeorphysm invariant and unitary massive supergravity \cite{TsujikawaModestoIR} is defied by the following nonlocal action,
\be 
S_{\rm m} \! = \frac{1}{\kappa^2} \! \int \! d^8 z  E^{-1} \left(-3
+\frac{1}{2}  \, {R} \, %\gamma_{\mathcal{R}}(
\frac{m^2}{\Delta_{{R}}^2}  \, \bar{ {R}} 
+ G^\mu \,  %\gamma_{G}( \Delta_{G} ) 
\frac{m^2}{\Delta_{{G}}^2} 
\, G_{\mu} \right) % \nonumber \\
 \, 
\label{MassiveSNL1}
\ee
that is obtained simply replacing $e^{\rm H} \rightarrow (-\Delta_{R,G}+m^2)/(-\Delta_{R,G})$ in (\ref{massiveForm}) and $m$ 
is the % graviton %(actually 
gravitational multiplet mass. It is easy to derive the number of Bosonic and Fermonic components using the propagators for the components fields (\ref{SPf}).  
In a previous work \cite{TsujikawaModestoIR} it has been explicitly proved the tree-level unitarity (optical theorem) of the purely Bosonic theory.
To probe the tree-level unitarity we coupled the propagator to the most general external conserved energy tensor $T_{\mu\nu}$, and afterward we examined the tensor-tensor amplitude at the 
massive pole. The transition amplitude in momentum space turned out to be positive at the pole \cite{HigherDG}, 
namely 
 \be 
 2 \, {\rm Im} \left\{  T(k)^{\mu\nu} \mathcal{O}^{-1}_{\mu\nu, \rho \sigma} T(k)^{\rho \sigma} \right\} = 2  \pi \, {\rm Res} \left\{  T(k)^{\mu\nu} \mathcal{O}^{-1}_{\mu\nu, \rho \sigma} T(k)^{\rho \sigma} \right\} \big|_{k^2 =-m^2}> 0.
 \label{Amplitude}
 \ee
 where $T^{\mu\nu}(k)$ is the conserved energy tensor in the Fourier space and $\mathcal{O}^{-1}_{\mu\nu, \rho \sigma}$ is the propagator in the momentum space. Notice, that for the theory 
(\ref {MassiveSNL1}) the tensorial structure of the massive graviton propagator is $P^{(2)} - P^{(0)}/2$, 
the same of the massless one. Nevertheless, the optical theorem is satisfied and 
the generalization of the tree-level unitarity proof to supergravity is straightforward. 

\subsection{Nonlocal Pauli-Fiertz Supergravity}\label{Pauli}
In this section we propose a general covariant and supersymmetric action whose Bosonic sector reduces to the linearized Pauli-Fiertz (PF) action for the massive spin two graviton. 
In constructing the supersymmetric theory we follow again the analogy with the Bosonic theory explicated in the previous sections.
%Therefore, 

Let us start with the nonlocal general covariant action for the PF massive gravity, % reads:
\be
&& \mathcal{L}_{\rm PF} = -  2 \kappa^{-2} \sqrt{|g|}\Big[ {\bf R} +
 {\bf C} %\mu \nu \rho \sigma}
  \gamma_{\rm C} (-\Box_{\Lambda}) {\bf C}
  + {\bf R}  \gamma_{\rm S}(-\Box_{\Lambda}) {\bf R }
  %C^{\mu \nu \rho \sigma} 
  %
 \Big]  \,  ,
 \nonumber 
 \\
&&
\gamma_{\rm C} =  \frac{1}{2}  \frac{e^{{\rm H}_2} -1}{\Box} \, , \quad 
\gamma_{\rm S} = - \frac{1}{6}  \frac{e^{{\rm H}_0} -1}{\Box} \, .
\label{TWeyl2}
\ee
where now we make the following replacement, 
\be
&& e^{{\rm H}_2} \,\,\rightarrow \,\, e^{{\rm H}_2} \left(   \frac{- \Box +m_1^2}{-\Box}  \right) \nonumber \\
&& e^{{\rm H}_0} \,\,\rightarrow \,\, - e^{{\rm H}_0} \left(  \frac{- \alpha \Box + m_2^2}{-\Box}  \right) 
\ee
It is easy to compute the propagator using the results published in \cite{modesto}. In particular the gauge invariant part of the two points function in momentum space now displays $5+1$ degrees of freedom,
% as evident from the tree-level scattering amplitude,
\be
\mathcal{O}^{-1}_{\mu\nu, \rho \sigma} = e^{-{\rm  H}_2}  \frac{P^{(2)}}{k^2 + m_1^2} 
\label{propagatorM} 
+ e^{- {\rm H}_0} \frac{P^{(0)}}{  2 ( \alpha k^2+m_2^2)  } \,  ,
\ee
where $\alpha$ is a real parameter. 
Since we are here interesting in infrared modifications to gravity, we fix ${\rm H}_2= {\rm H}_0 =0$ in the formula above. 
Notice that for $\alpha =0$ the zero mode does not propagate and the spectrum consist only of the 
massive graviton. In particular the amplitude (\ref{Amplitude}) reads,
\be
  T^{\mu\nu} \mathcal{O}^{-1}_{\mu\nu, \rho \sigma} T^{\rho \sigma}  =  
  \frac{T_{\mu\nu}  T^{\mu\nu} - \frac{1}{3}  T^2}{k^2+m_1^2} . 
\ee

The nonlocal PF gravity is easily obtained in the superspace by making the following replacements in the form factors (\ref{gammaWR}),
\be
&&  e^{\rm {\rm H}_2(\Delta_W)} \,\, \rightarrow \,\, \frac{- \Delta_{W}+m_1^2 }{-\Delta_{W} } \, 
 , \nonumber \\ % \quad
 && 
 e^{\rm {\rm H}_0(\Delta_W)} \,\, \rightarrow \,\, -  \frac{- \alpha \Delta_{R}+m_2^2 }{-\Delta_{R} } \,  , \\
%\gamma_{ R} = - \frac{3}{\kappa^2} \frac{e^{\rm H_0(\Delta_R)} -1}{\Delta_R} -4  \gamma_G \, . 
&& 
{\rm and} \quad 
\gamma_W = \frac{ 1}{\kappa^2} \frac{m_1^2}{\Delta_W} \, , \quad
\gamma_R = \frac{ - 3 m_2^2 + 3 \Delta_R( 1 + \alpha) }{ \kappa^2 \Delta_R^2} \,   ,
%
% - \frac{3}{\kappa^2} \frac{e^{\rm H_0(\Delta_R)} -1}{\Delta_R} -4  \gamma_G \, . 
\label{gammaWRmassive}
\ee
where for the sake of simplicity we assumed $\gamma_G=0$. Finally, the PF massive supergravity reads:

\be
S_{\rm NL-PF} 
= \frac{1}{\kappa^2}  \! \int \! d^8 z E^{-1} \left(- 3 %\frac{3}{\kappa^{2}}  
- 3 {R} \,  \frac{  m_2^2 -  \Delta_R( 1 + \alpha) }{  \Delta_R^2}
%\gamma_{{R}}(\Delta_{{R}} )
 \, 
\bar{{R}} 
%+ G^{\alpha\dot\alpha} \,  \gamma_{G}( \Delta_{G} ) \, G_{\alpha\dot\alpha} 
%\right)  + \int \! d^8 z 
%+ \frac{E^{-1}}{ \mathcal{R}}  
+ {R}^{-1}
W_{\alpha \beta \gamma} \, 
 \frac{m_1^2}{\Delta_W}
%\gamma_{W}( \Delta_{W} )
\, 
W^{\alpha \beta \gamma}  \right)   .  %}
\!\!\! 
\label{PFmS}
\ee
We can now read out the spectrum from the propagators for the component fields.
Again we fix $\alpha =0$ and the propagators simplify to:
\be
&& \langle h \,h \rangle \propto %= 8 
\frac{ P^2}{ \Box - m_1^2}   +  %\frac{1}{ \Box - m^2 } 
\frac{P^{0,s}}{2 m_2^2}   \, , 
 \, \quad
 \langle \psi \, \psi \rangle \propto %= - 2
  \frac{ \not\!\partial}{ \Box - m_1^2}  P^{3/2} +  %\frac{ \not\!\partial}{ \Box - m^2}  
 \frac{2}{m_2^2} P^{1/2}   \, , \nonumber  \\
&& 
\langle A \, A \rangle  \propto 
\frac{\Box}{\Box - m_1^2}  P^1 +
%\frac{\Box }{\Box - m^2}    
  \frac{\Box}{m_2^2}  P^{0}   %\gamma^{-1} 
\, ,  \, \quad 
\langle B \, \bar{B} \rangle \propto 
 \frac{\Box}{m_2^2} 
\label{SPmassive} \,.
\ee
The number of degrees of freedom is: $5$ for the graviton, $8$ for the massive gravitino, 
$3$ for the vector. For $\alpha >0$ the spectrum extends to include the massive multiplet 
$(1/2, 0^+, 0^-)$ of the Starobinsky's supergravity in the previous section.

%%%%%%%%%%%%%%%
%linearized propagator
%%%%%%%%%%%%%%%
\section{Quantum nonlocal supergravity: super-renormalizability and finiteness} %Power counting

In this section we show the quantum super-renormalizability of the weakly nonlocal supergravity. Using the notations of section \ref{LinearSUGRA} and in particular the rescaled dimensionless superfield $H_{\alpha \dot{\alpha}}$,
we can easily derive the structure of the divergences of the quantum theory. 
A crucial property of the theory in addition to weak nonlocality is ``quasi polinomiality", i.e.
the form factors appearing in the action must be polynomial for large values of their argument.
This behavior guarantees locality of the counterterms. A theory based on form factors with exponential asymptotic behavior, i.e.
$\gamma(\Box) = \exp (- \Box)^n$ ($n \in \mathbb{N}^+$), poses additional challenges that are still debated in the literature.  
Once we express the form factor as the exponential of an entire function $\exp {\rm  H} (z)$, 
an example of entire function ${\rm H}(z)$ is given in \cite{Kuzmin, Tombo}, 
%%%
\be
%&& \hspace{-1cm} 
{\rm H}(z) =  \frac{1}{2} \left[ \gamma_E + 
\Gamma \left(0, p_{\gamma+1}^{2}(z) \right) \right] + \log [ p_{\gamma+1}(z) ] 
\label{H}   \equiv \sum_{n =1}^{+ \infty} ( -1 )^{n-1} \, \frac{p_{\gamma+1}(z)^{2 n}}{2n \, n!} \,  , 
%\quad {\rm Re}( p_{\gamma+1}^{2}(z) ) > 0  \, , 
\label{FormFactor}
\ee
where $p_{\gamma+1}(z)$ is a real polynomial of degree $\gamma+1$ 
(${\rm Re}( p_{\gamma+1}^{2}(z) ) > 0$), 
$\gamma_E$ %\approx 0.577216$ 
is the Euler's constant and  
$\Gamma(a,z)$ %= \int_z^{+ \infty} t^{a -1} e^{-t} \rm{d} t$ 
is the incomplete gamma function.  
If we choose $p_{\gamma+1}(z) = z^{\gamma+1}$, %%%%%
 the $\Theta$ angle defining the conical region around the real axis in which we have asymptotic polynomial behavior is % the cone $C$,  
 $\Theta = \pi/4 (\gamma+1)$. %\gamma+4)$. 
The crucial property for having only local counter terms mentioned above reads as follows,
\be 
\hspace{-0.4cm}
e^{\frac{1}{2} \left[ \Gamma \left(0, p(z)^2 \right)+\gamma_E  + \log \left( p(z)^2 \right) \right] }=
e^{\frac{\gamma_E}{2}}
%\left|
\sqrt{ p(z)^2} % \right|  %\Bigg\{ 1+
%\label{Vlimit1} 
\left\{ 
1+ \left[ \frac{e^{-p(z)^2}}{2 p(z)^2} \left(  1 %\frac{1}{2 p(z)^2} %- \frac{1}{2 p(z)^4} 
+ O \left(   \frac{1}{p(z)^2} \! \right)   \! \right) + O \left(e^{-2p(z)^2} \right)  \right] \right\}  \!  , 
\ee
by which $\exp {\rm  H} (z)\approx z^{\gamma+1}$ for large $z$.
First we describe the general structure of the supergraph diagrams appearing in the loop expansion. We have already determined the quadratic action \eqref{SNL2}, from which one can read the propagators for $H_{\alpha \dot{\alpha}}$, $\sigma$ and $\overline{\sigma}$. This requires the choice of  a gauge, the most convenient one being
\begin{equation}
D^{\alpha}H_{\alpha \dot{\alpha}}=\overline{D}^{\dot\alpha}H_{\alpha \dot{\alpha}}=0\,,
\end{equation}
which is such that \eqref{SNL2}  takes the diagonal form
\begin{equation}
S_{\rm NL}^{(2)}=\int d^{8}z\left[H^{\alpha\dot\alpha}\square h_{\frac{3}{2}}(\square)H _{\alpha\dot\alpha}+9\sigma h_{0}(\square)\overline{\sigma}\right] \, .
\label{eq:DiagNLsugra}
\end{equation}
The internal lines can be of four kinds: $HH$, $\sigma\overline{\sigma}$, $\sigma\sigma$ and $\overline{\sigma}\overline{\sigma}$, but the latter two are vanishing for massless chiral fields, which is our case too. The two non-null superpropagators are
\be
&& P_{HH}\,_{\alpha \beta}\,^{\dot{\alpha} \dot{\beta} } (x - x^{\prime})
=\int  \frac{d^4 p}{(2\pi)^4}  e^{- i p(x - x^{\prime}) } \frac{-1}{2p^2}\, h_{\frac{3}{2}}^{-1}(-p^2/\Lambda^2)\delta_{\alpha}{}^{\dot\alpha}\delta_{\beta}{}^{\dot\beta} \delta^{(4)}(\theta - \theta^{\prime})\,,\label{eq:SupPH}\\
&& P_{\sigma\overline{\sigma}} (x - x^{\prime})
 = \int \frac{d^4 p}{(2\pi)^4}e^{- i p(x - x^{\prime}) }\frac{-i}{9p^2}h_{0}^{-1}(-p^2/\Lambda^2)\delta^{(4)}(\theta - \theta^{\prime}) \,.\label{eq:SupPSigma}
\ee
If we choose 
\be
h_{\frac{3}{2}}(z)= \frac{1}{2} e^{ H_{2}(z)}  \quad  \mbox{and} \quad  
h_{0}(z) =- \frac{1}{3} e^{H_{0}(z)} 
\ee
 with $H_{2}(z)$ and $H_{0}(z)$ entire functions of the kind described in \eqref{FormFactor}, both propagators at high energy scale as: 
 \be
 \frac{1}{k^{(2 \gamma +4)}} \, . 
 \ee
 % at high energy.  
%
To get the interaction vertices we should expand action \eqref{SNL2} to higher orders, which is a quite technically involved task because it is necessary to consider the contributions given by the expansion of the form factors. Luckily, for the purpose of power counting analysis, the detailed structure of vertices is not necessary. Each interaction vertex $i$ is characterized by the numbers $n_{i\,H}$, $n_{i\,\sigma}$ and $n_{i\,\overline{\sigma}}$ of superfields $H_{\alpha \dot{\alpha}}$, $\sigma$ and $\overline{\sigma}$ respectively and by the number $d_i$ of spinor derivatives acting on these superfields in the UV regime. We need observe that the weak superfield expansion is such that for each $H_{\alpha \dot{\alpha}}$ two spinor covariant derivatives will show up in the vertices, whereas 
$\sigma$ and $\overline{\sigma}$ will not bring any derivatives. Another crucial observation is that, once we fix the number of  superfields 
\be
n_i=n_{i\,H}+n_{i\,\sigma}+n_{i\,\overline{\sigma}} \, , 
\ee
the number of derivatives $d_i$ is given by the formula:
\be
d_i=2n_{i\,H}+{4\gamma+4} \,  ,
\ee
irrespective of how many of these fields come from the expansion of the form factors. So different vertices, in particular with a different dependence on the form factors, are characterized by the same structure in the UV regime, which is what is ultimately relevant as far as power counting is concerned. In a bookkeeping notation, such that we ignore indices and combinatorial factors, we can therefore represent the generic interaction term as
\be
 &&
S_{{\rm int}, i } =  %\kappa^{-2} 
\kappa^{n_i-2}\Lambda^{-2( \gamma+1) }
 \int d^8 z   \, \underbrace{ ({D} \bar{{D}} H)  ({D} \bar{{D}} H)
\dots  ({D} \bar{{D}} H)}_{n_{i\,H}}
 \Box^{\gamma+1}   \underbrace{\sigma\ldots\sigma }_{n_{i\,\sigma}}\underbrace{\overline{\sigma}\ldots\overline{\sigma} }_{n_{i\,\overline{\sigma}}}\,.%{\rm h.c.} .
  \label{eq:nG}
\ee
For the sake of simplicity, we shall assume $\kappa\equiv\Lambda^{-1}$ in the following. As each vertex contains an integral over $d^4\theta$, an arbitrary $L$-loop supergraph $G$, with $P$ propagators, $V$ vertices, and $E$ external lines, has the form
\begin{equation}
\int d^4 p_1 \ldots d^4 p_L d^4\theta_1 \ldots d^4\theta_V \left[\ldots\right]\,,
\label{eq:supergraph}
\end{equation}
where the square brackets include the above number of propagators and some definite number of $D$-factors associated with vertices. We remind that: $[H]=0$ and $[\sigma]= [\bar{\sigma}] =1$. 
The momentum integrals in \eqref{eq:supergraph} contribute the quantity $4L$ to the superficial degree of divergence $\omega(G)$. Taking into account the explicit dependence of  the propagators \eqref{eq:SupPH} and \eqref{eq:SupPSigma} on momenta, we find their contribution to $\omega(G)$ is $- (2 \gamma +4)P$. Noticing that the vertex \eqref{eq:nG} contains the factor $k^{2\gamma+2}$, we can find the quantity
\begin{equation}
4L- (2 \gamma +4)P+(2\gamma+2)V\,,
\end{equation}
contributing to the degree of divergence $\omega(G)$. Owing to the well-known topological relation
\begin{equation}
V + L -P=1\,,
\end{equation}
this quantity can be recast in the following form, %rewritten as
\begin{equation}
2-2\gamma(L-1)+2(L-P)\,.
\label{eq:momenta}
\end{equation}
However, the final momentum dimension of the supergraph $G$ receives a contribution also from the $D$-factors, which depend on the momentum too. The superfield Feynman rules are such that each vertex  without external lines includes $2n_i$ $D$-factors whereas we have to subtract $2$ $D$-factors for each external line. So, if $V_i$ is the number of vertices of type $i$ appearing in the supergraph $G$ we find the total number of $D$-factors depending on the internal momenta in expression \eqref{eq:supergraph} is given by
\begin{equation}
2\sum_i V_i n_i -2E\,.
\label{eq:topo}
\end{equation}
Now we should remind %remember 
%that even in the case of such supergravity theories as the ones described here a 
that a non-renormalization theorem can be proven also for the weakly nonlocal supergravity. % by which 
Therefore, each supergraph forming the effective action can be represented as a single integral over $d^4\theta$. This means that in \eqref{eq:supergraph} $V-1$ $\theta$-integrals can be taken explicitly, since each internal line contains a Grassmann delta-function $\delta^4(\theta_i-\theta_j)$. So in the end the number of remaining $\delta$-functions equals 
\begin{equation}
P-V+1 = L\,,
\end{equation}
where the topological relation \eqref{eq:topo} has been used again. Since $\delta^4 (\theta-\theta)=0$ the expression \eqref{eq:supergraph}, rephrased as a single $\theta$-integral of an integrand containing $L$ Grassmann delta-functions, can be non-zero only if $4L$ $D$-factors cancel all the $\delta$-functions by means of relations like 
\be
{D}^2 \overline{D}^2\delta^4(\theta-\theta^\prime)=16 \, . 
\ee
So, thanks to the non-renormalization theorem, we have the following number of $D$-factors depending on the internal momenta
\begin{equation}
2\sum_i V_i n_i -2E-4L\,.
\end{equation}
Some of these $D$-factors can be transferred to the external lines integrating by parts whereas the remaining ones must be converted into momenta by the law $\{{D}, \overline{D}\}\sim p$. Let us now consider the maximally divergent case when all the $D$-factors are converted into momenta. Notice that this case can be realized only if the supergraph $G$ contains an equal number of internal $\sigma$- and $\overline{\sigma}$-superfields, namely 
\be
\sum_i V_i n_{i\sigma}-E_{\sigma}=\sum_i V_i n_{i\overline{\sigma}}-E_{\overline{\sigma}} \, . 
\ee
In general the number of convertible $D$-factors is given by
\begin{equation}
2\sum_i V_i n_{iH}+4\left[\frac{\sum_i V_i n_{i\sigma}-E_{\sigma}+\sum_i V_i n_{i\overline{\sigma}}-E_{\overline{\sigma}}}{2}\right]-2E_H-4L\,,
\end{equation}
producing the following maximal number of internal momenta 
\begin{equation}
\sum_i V_i n_{iH}+2\left[\frac{\sum_i V_i n_{i\sigma}-E_{\sigma}+\sum_i V_i n_{i\overline{\sigma}}-E_{\overline{\sigma}}}{2}\right]-E_H-2L\,.
\end{equation}
This number is greatest for supergraphs such that 
\be
\sum_i V_i n_{i\sigma}-E_{\sigma}=\sum_i V_i n_{i\overline{\sigma}}-E_{\overline{\sigma}}
\ee
(e.g. such that $\sum_i V_i n_{i\sigma}=\sum_i V_i n_{i\overline{\sigma}}=0$), in which case it can be simply expressed as
\begin{equation}
\sum_i V_i n_{i}-E-2L\,.
\label{eq:Dmomenta}
\end{equation}
Then, the maximal superficial degree of divergence is given by the sum of \eqref{eq:momenta} and \eqref{eq:Dmomenta}
\begin{equation}
\omega_{\rm max}(G)=2-2\gamma(L-1)+\sum_i V_i n_{i}-E-2P\,,
\end{equation}
which, owing to the relation $E+2P=\sum_i V_i n_{i}$ (or $\sum_i V_i n_{i} - 2L= E-2$), simplifies to
\begin{equation}
\omega_{\rm max}(G)=2-2\gamma(L-1)\,.
\end{equation}
Therefore the supergraph $G$ can be sketchily represented as
\be
&&
% \delta^4(K) \,  \,   \int  ( d^4 \theta )
\hspace{-1cm}
\Lambda^{ 2 \gamma  (L-1)} \left( \Lambda_{\rm cut-off} \right)^{\omega(G)}  \int d^4x d^4 \theta \times \nonumber\\
&&\hspace{-1cm}
 \times  \underbrace{ (\Lambda^{-1} {D} \bar{{D}} H)  (\Lambda^{-1}  {D} \bar{{D}} H)
\dots  (\Lambda^{-1 } {D}  \bar{{D}} H)}_{E_{H}}\,\partial^{\frac{N_D}2}\underbrace{(\Lambda^{-1} \sigma)\ldots(\Lambda^{-1} \sigma) }_{E_{\sigma}}\underbrace{(\Lambda^{-1} \overline{\sigma})\ldots(\Lambda^{-1} \overline{\sigma}) }_{E_{\overline{\sigma}}} \, ,%\nonumber \\ %\quad 
%&& 
%\omega(G) \equiv 2 - 2 \gamma  (L - 1) \, .
\label{PC}
\ee
where 
\be
\omega(G)=2 - 2 \gamma  (L - 1) - {N_D}/{2} \, , 
\ee
 and $N_D$ the number of $D$-factors transferred to the external lines.
The structure of the perturbative counterterms needed to cancel the divergences in \eqref{PC} is determined by the superficial degree of divergence $\omega(G)$, the non-renormalization theorem and the fact that counterterms are local functionals in $x$-space. Assuming it is possible to choose a regularization scheme preserving supersymmetry, the generic counterterm $\Delta S$ has the form of a superspace integral $d^8 z$
\be
\Delta S = \Lambda^{ 2 \gamma  (L-1)}  \int d^8 z \, \Delta\mathcal{L}_{\rm ct} \left(\Lambda^{-1}\sigma, \Lambda^{-1}\overline{\sigma},\Lambda^{-1} {D} \bar{{D}} H, \ldots \right) \, , \quad [ \Delta\mathcal{L}_{\rm ct} ] = {\omega_{\rm max}(G)} . 
\ee
%\be 
%&&
%\omega(G) =  D - 2 \gamma  (L - 1)    \, , \,\,\,\, 
%\label{even}
%\ee
where 
\be
\Delta\mathcal{L}_{\rm ct} \left(\Lambda^{-1}\sigma, \Lambda^{-1}\overline{\sigma},\Lambda^{-1} {D} \bar{{D}} H, \ldots \right)
\ee
is a function of the basic superfields and their derivatives up to order ${\omega_{\rm max}(G)}$. One can actually determine the general structure of counterterms by the background field method, implying the effective action must be invariant under the classical super-diffeomorfism transformations. So the local divergent part of the effective action is
\be
\Delta S =\Lambda^{ 2 \gamma  (L-1)}  \int d^8 z E^{-1}  \Delta\mathcal{L}_{\rm ct}\,,
\ee 
where $ \Delta\mathcal{L}_{\rm ct}$ is a product of factors of supervierbein and connections. To insure covariance these factors must arrange themselves in a form which contains at most one non covariant object times a covariant object which satisfies a Bianchi identity.

If $\gamma > 1$, only 1-loop divergences survive.  This bound is less restrictive then the one found for purely Bosonic super-renormalizable gravity, i.e. $\gamma>2$ coming from the maximal superficial degree of divergences $\omega_{\rm max}(G)=4-2\gamma(L-1)$.
Therefore, 
the theory is super-renormalizable \cite{Tombo,modesto,modestoLeslaw}
and only a finite number of counterterms has to be
included in the action, namely, as can be seen from \eqref{PC},  the terms with at most two derivatives. In particular, for the chosen monomial asymptotic behavior of the form factors \eqref{FormFactor} and adopting dimensional regularization scheme (i.e. in the absence of additional scales ), the only possible counterterms are the ones of momentum dimension two
%Heading back again to the superspace formulation, only the following operators %with dimension zero 
%are suitable as counter terms,  
\be
% \int \! d^8 z E^{-1} -\frac{3}{\kappa^{2}}  \, , \quad 
 && 
 \hspace{-0.3cm}
 \int \! d^8 z E^{-1}  {R}  \bar{{R}} \, ,\quad
\int \! d^8 z E^{-1}  G^{\alpha\dot\alpha} G_{\alpha\dot\alpha} \, , \quad
 \int \! d^8 z  \frac{E^{-1}}{R}
W_{\alpha \beta \gamma}   W^{\alpha \beta \gamma} + {\rm h.c.} \, , \quad \int \! d^8 z E^{-1}  \left( {R}^2 +  \bar{R}^2 \right) \, . \nonumber \\
&&
\label{superCounter}
\ee
Notice the last term but one can be obtained from a superspace integral of a local combination containing a supertorsion $W_{\alpha \beta \gamma}$ and therefore does not violate the non-renormalization theorem. 

In the case of non-monomial asymptotic behaviuor or if an explicit cut-off is used to regularize divergent supergraph, we can also have a counterterm of momentum dimension zero, namely the Einstein-Hilbert supergravity term
\be
  \int \! d^8 z E^{-1} \, , 
  \ee
  and counterterms of momentum dimension one, namely 
\be
\int \! d^8 z E^{-1}  ( {R}+\overline{R} ) \, , 
\ee
needed to cancel a linear divergence. This latter term is known to be a pathological one related to the possibility of getting terms non-polynomial in the curvature in the component action after elimination of the auxiliary fields.  %As for 
Finally, the counterterm for the cosmological constant should have the form 
\be
 \int \! d^8 z \frac{E^{-1}}{R}+ {\rm h.c.} \, . 
 \ee
However, such a cosmological constant term has momentum dimension $-1$ and can not be generated at quantum level. This is a nontrivial consequence of the non-renormalization theorem enforcing the counterterms to have the structure of full superspace integrals.
Moreover, one of the divergences \eqref{superCounter} can be removed by means of the super-Gauss-Bonnet identity 
(\ref{SGB}). 

%It will be shown in the next section the remaining divergences don't actually affect gauge-independent on-Einsten shell quantities, in particular the S-matrix and therefore their physical relevance is quite limited. In any case one can ask the question whether it is possible to get rid of these one-loop divergences even off-shell. 

The strategy we pursue here to eliminate these one-loop divergences is to add some gauge invariant interaction terms providing analogous counterterms with the opposite sign. This procedure is fully consistent because in an higher derivative theory operators of 
dimension greater then two are not subject to infinite renormalizations and, therefore, their coupling do not run. 
%vertex renormalizations are finite and so no additional divergence needs be regularized. 
It is clear that both the divergences and the new vertices %to be introduced to kill them 
are strongly constrained by the requirement of general covariance. The best formalism to keep covariance explicit is the background field method whereby %a proper gauge can be chosen such that
the quadratic part of the supergravity action, together with a proper gauge fixing, takes the form
\begin{equation}
S_{\rm NL}^{(2)}=\int d^{8}z\left[\mathbf{H}^{\alpha\dot\alpha}\mathcal{H}_{\alpha\dot\alpha\,,\beta\dot\beta}\mathbf{H}^{\beta\dot\beta}+9\mathbf{\sigma} \mathcal{S}\,\overline{\mathbf{\sigma}}+\ldots\right] \, ,
\label{eq:BackgroundSNL}
\end{equation}
where
$\mathcal{H}^{\alpha\dot\alpha\,,\beta\dot\beta}$ and $\mathcal{S}$ are derivative operators which are determined by the second variation of \eqref{SNL}. Of course, in the background-quantum splitting formalism with a generic curved background also the first and second variations of the form-factors are to be taken into account and $\mathcal{H}^{\alpha\dot\alpha\,,\beta\dot\beta}$ and $\mathcal{S}$ will have a very complicated dependence on background covariant derivatives and derivatives of the supercurvatures. Luckily, we do not need to know the explicit shape of \eqref{eq:BackgroundSNL} to determine the kind of divergent terms appearing in the quantum effective action. We can just consider the asymptotic behavior of $\mathcal{H}_{\alpha\dot\alpha\,,\beta\dot\beta}$ and $\mathcal{S}$ in the UV.
Actually (\ref{eq:BackgroundSNL})
is the generalization of the quadratic action \eqref{eq:DiagNLsugra} for a generic curved background. Notice that $\mathbf{H}^{\alpha\dot\alpha}$ is now a real tensor superfield defined by the gauge choice $\mathbf{H}=\mathbf{H}^a\mathcal{D}_a$ and $\mathbf{\sigma}$ a covariantly chiral superfield related to the covariantly chiral compensator $\mathbf{\varphi}=e^{\mathcal{\sigma}}$, $\overline{\mathcal{D}}_{\dot\alpha}\mathbf{\varphi}=0$. They have the following quantum  gauge transformations
\be
\delta \mathbf{H}^{\alpha\dot\alpha}&=&\overline{\mathcal{D}}^{\dot\alpha}L^\alpha-\mathcal{D}^\alpha L^{\dot\alpha}+O(\mathbf{H})+O(\mathbf{\sigma})\,,\\
\delta \mathbf{\varphi}^3&=&\frac14\left(\overline{\mathcal{D}}^2-4R\right)\mathcal{D}_\alpha(L^\alpha\mathbf{\varphi}^3) \, .
\ee
Such gauge symmetry can be fixed in such a way as to get rid of mixed quadratic terms that can generically appear, leading to the diagonal form of \eqref{eq:BackgroundSNL}. The dots stand for the ghosts that appear because of the Faddeev-Popov procedure. These do not include only the usual Faddeev-Popov ghosts, but also hidden ghosts related to the fact that the gauge-fixing conditions can be subject to constraints and the extra Nielsen-Kallosh ghosts are necessary to correctly normalize the gauge averaging. A detailed analysis of the quantization procedure and of the related ghosts is beyond the scope of this paper. For the time being we assume that, as it is usual in higher derivative theories, the gauge-fixing weighting functions can be chosen so that the one-loop divergences are explicitly gauge invariant and do not involve the ghosts. 

\begin{comment}
 which is dominated by the following terms
\be
\mathcal{H}_{\alpha\dot\alpha\,,\beta\dot\beta}&=&\omega_1 \varepsilon_{\alpha\delta}\varepsilon_{\dot\alpha\dot\delta}\mathbf{\Box}^{\gamma+2}
\ee
\end{comment}

%%
%The answer to this question is quite simple. 
To make the theory finite we now explicitly introduce the announced 
extra super-symmetric operators that contribute to the beta functions, but do not get quantum renormalizations. 
These operators are the analog of the ``killer operators" introduced to make finite the Bosonic theory.
\cite{modestoLeslaw}. 
Three candidate super-killers 
can be easily defined in the superspace, 
\be
%&& s_{K_{R }} \! \int \! d^8 z E^{-1}  {R} \left(\overline{\mathcal{D}}^2-4R\right)\bar{{R}} ( \Delta_{\mathcal{R}})^{\gamma -2} {{R}}  \left(\overline{\mathcal{D}}^2-4R\right)\bar{{R}} 
%+{\rm h.c.}\, ,  \label{SK0} \\
&& s_{K_{R }} \! \int \! d^8 z E^{-1}  {R} \bar{{R}} \, 
( \Delta_{\mathcal{R}})^{\gamma -1} 
%\left({\mathcal{D}}^2-4 \bar{R} \right)R  \left(\overline{\mathcal{D}}^2-4R\right)
R \bar{{R}} 
+{\rm h.c.}\, ,  \quad 
\tilde{s}_{K_{R }} \! \int \! d^8 z E^{-1}  R R \, 
( \Delta_{\mathcal{R}})^{\gamma -1} 
%\left({\mathcal{D}}^2-4 \bar{R} \right)R  \left(\overline{\mathcal{D}}^2-4R\right)
R R  +{\rm h.c.}\, , 
 \label{SK1} \\
%&& s_{K_{R }} \! \int \! d^8 z E^{-1}  {R} \bar{{R}} \, 
%\left(\overline{\mathcal{D}}^2-4R\right)( \Delta_{\mathcal{R}})^{\gamma -2} 
%\left({\mathcal{D}}^2-4 \bar{R} \right)
%R  \bar{{R}} 
%+{\rm h.c.}\, ,  \label{SK1} \\
&& s_{K_{G}} \! \int \! d^8 z E^{-1} G_{\alpha\dot\alpha} G^{\alpha\dot\alpha} \,  
( \Delta_{{\mathcal{R}}})^{\gamma -1}  
%\left({\mathcal{D}}^2-4 \bar{R} \right)R  \left(\overline{\mathcal{D}}^2 - 4 R \right) \bar{{R}} 
G_{\beta\dot\beta} 
G^{\beta\dot\beta} 
+{\rm h.c.}\, ,
\label{SK2}\\
%%%
%&& s_{K_{G}} \! \int \! d^8 z E^{-1} G_{\alpha\dot\alpha} G^{\alpha\dot\alpha} \,  
%( \Delta_{{G}})^{\gamma -2}\, 
 %\left(\overline{\mathcal{D}}^2-4R\right)
%G_{\beta\dot\beta} 
%\left(\overline{\mathcal{D}}^2-4R\right)
%G^{\beta\dot\beta} +{\rm h.c.}\,, 
%\label{SK2}\\
&& s_{K_{W}} \! \int \! d^8 z  \frac{E^{-1}}{R}
W_{\alpha \beta \gamma} W^{\alpha \beta \gamma}\, ( \Delta_{{\mathcal{R}}})^{\gamma - 1}
%\left({\mathcal{D}}^2-4 \bar{R} \right)R  \left(\overline{\mathcal{D}}^2 - 4 R \right) \bar{{R}} 
W_{\rho \delta \tau} W^{\rho \delta \tau} 
+{\rm h.c.} \,  . 
\label{SK3} 
%%%%
%&& s_{K_{W}} \! \int \! d^8 z  \frac{E^{-1}}{R}
%W_{\alpha \beta \gamma} \left({\mathcal{D}}^2-4\overline{R}\right)W^{\alpha \beta \gamma}\, ( \Delta_{W})^{\gamma - 2} \,   W_{\rho \delta \tau}
%\left({\mathcal{D}}^2-4\overline{R}\right)W^{\rho \delta \tau}  +{\rm h.c.}\,  . 
%\label{SK3} 
\ee
The above operators can only give linear contributions to the one loop beta functions
for exactly the same reason recently discovered in the purely Bosonic theory \cite{modestoLeslaw}.
Indeed, applying the super-background field method, the second 
variations of the super-killers (\ref{SK1}), (\ref{SK2}), (\ref{SK3}) are proportional to the counter terms (\ref{superCounter}).
Therefore, the one-loop contributions to the beta functions will be linear in their front coefficients
and it is always possible to choose two out of the three parameters 
$s_{K_{\mathcal{R} }},s_{K_{G}},s_{K_{W}}$ to make zero the beta functions ending with a finite theory.
Of course this is the most expensive finite  supergravity. It is indeed possible that the theory here presented, 
or an extension to $N=2$ supersymmetry, is already finite without need of extra higher curvature operators (\ref{SK3}). 
Only an explicit computation of the one loop beta functions  will provide the final answer.

So far we showed that the weakly nonlocal supergravity is finite at quantum level in dimensional regularization scheme.  Now we investigate the possibility to cancel not only logarithmic divergences, but also the linear and quadratic one. This analysis requires to implement the cut-off regularization scheme. 
%%%%
%%%%%
\begin{comment}
Let us introduce here two more killer operators cubic in the superspace curvatures,
%%%
\be
 && s^{(a)}_{K_{E}} \! 
\int \! d^8 z E^{-1}  {R} ( \Delta_{\mathcal{R}})^{\gamma -1}  
\left({\mathcal{D}}^2-4 \bar{R} \right){R} 
\left(\overline{\mathcal{D}}^2-4R\right)\bar{{R}}+{\rm h.c.}\, , \label{E1} \\
%&& s^{(b)}_{K_{E}} \! 
%\int \! d^8 z E^{-1} \mathcal{D}^\alpha \overline{\mathcal{D}}^{\dot\alpha}G_{\alpha\dot\alpha}( \Delta_{{G}})^{\gamma -1}\,  G_{\alpha\dot\alpha} G^{\alpha\dot\alpha} +{\rm h.c.}\, . 
&& s^{(b)}_{K_{E}} \! 
\int \! d^8 z E^{-1}  R \, 
( \Delta_{{G}})^{\gamma -1 } 
 \left({\mathcal{D}}^2-4 \bar{R} \right)
G_{\beta\dot\beta} 
\left(\overline{\mathcal{D}}^2-4R\right)
G^{\beta\dot\beta} +{\rm h.c.}\, . 
%\mathcal{D}^\alpha \overline{\mathcal{D}}^{\dot\alpha}G_{\alpha\dot\alpha}( \Delta_{{G}})^{\gamma -1}\,  G_{\alpha\dot\alpha} G^{\alpha\dot\alpha} +{\rm h.c.}\, . 
\label{E2}
\ee
\end{comment}
%%%
In presence of a cut-off one can expect to generate extra divergences proportional to the following 
operators, 
\be
\int d^8 z E^{-1} \, , \quad  \int d^8 z E^{-1} (R + \bar{R}) ,
\label{Extra}
\ee 
regardless of the monomial asymptotic behaviour of the form factor/s referred at the beginning of this section. The beta functions have dimension $[ \beta_{ E^{-1}}] = 2$, 
$[ \beta_{E^{-1} (R + \bar{R}) }] = 1$. 
However, the second operator in (\ref{Extra}) has dimension one, 
while the divergent integrals should be linear in the cut-off. This is not possible in $D=4$
because the asymptotic behaviour of the entire functions is polynomial and all the Lorentz invariant integrals have the following structure,
\be
\int d^4 p \frac{p^{2 m}}{(p^2+C)^{2 n}} \, , \quad n,m \in \mathbb{N} \, ,
\ee
where $C$ is a function of the external moments. 

The first Einstein-Hilbert superspace operator in (\ref{Extra}) is the last divergence we have handle with to achieve 
finiteness in cut-off regularization scheme. 
%
%The beta functions have dimension $[ \beta_{ E^{-1}}] = 2$, $[ \beta_{E^{-1} (R + \bar{R}) }] = 1$. 
%The killer operators (\ref{E1}) and (\ref{E2}) give both a contribution 
%to the beta function 
%$\beta_{E^{-1}( R + \bar{R})}$ 
%linear in the front coefficients $s^{(a)}_{K_{E}}$ and/or $s^{(b)}_{K_{E}}$.
%These coefficients do not run and one out of the two can be fixed to make zero 
%the beta function for the second operator in (\ref{Extra}). 
%The only divergence that still survives in cut-off regularization scheme is proportional to the 
%Einstein supergravity operator (the first one in (\ref{Extra}).) 
In the Bosonic theory we can rid out the quadratic divergence 
introducing a Bosonic killer operator cubic in the curvature, namely $\int \mathscr{R}^2 \Box^{\gamma - 1} \mathscr{R}$ \cite{FiniteEntang}.
Therefore, by analogy the last operator we need to introduce in the superspace to make finite the theory in the cut-off scheme must reduce to the above one when explicitly expressed in components. 
In the superspace two candidate operators are:
\be
 && s^{(a)}_{K_{E}} \! 
\int \! d^8 z E^{-1}  {R} \, ( \Delta_{\mathcal{R}})^{\gamma -1}  
\left(\overline{\mathcal{D}}^2-4R\right)
\left({\mathcal{D}}^2-4 \bar{R} \right){R} 
\, , \label{E1} \\
%&& s^{(b)}_{K_{E}} \! 
%\int \! d^8 z E^{-1} \mathcal{D}^\alpha \overline{\mathcal{D}}^{\dot\alpha}G_{\alpha\dot\alpha}( \Delta_{{G}})^{\gamma -1}\,  G_{\alpha\dot\alpha} G^{\alpha\dot\alpha} +{\rm h.c.}\, . 
&& s^{(b)}_{K_{E}} \! 
\int \! d^8 z E^{-1}  \bar{R}  \, 
( \Delta_{{\mathcal{R}}})^{\gamma -1 } 
 \left({\mathcal{D}}^2-4 \bar{R} \right) 
\left(\overline{\mathcal{D}}^2-4R\right)
\bar{R} \, .  
%\mathcal{D}^\alpha \overline{\mathcal{D}}^{\dot\alpha}G_{\alpha\dot\alpha}( \Delta_{{G}})^{\gamma -1}\,  G_{\alpha\dot\alpha} G^{\alpha\dot\alpha} +{\rm h.c.}\, . 
\label{E2}
\ee

In force of our analysis, we are moved to declare that the weakly nonlocal supergravity here proposed is finite not only in dimensional regularization, but most likely also in the cut-off regularization scheme.

\section{Scattering amplitudes}
The analysis of divergences we have presented so far has shown that by a mild relaxation of the assumption of locality, it is possible to define an action for quantum (super-)gravity exhibiting the same perturbative spectrum as ordinary two-derivatives (super-)gravity and at the same time improved UV behaviour. In particular, the two fundamental ingredients of the procedure adopted to get rid of the UV-divergences proliferating in second-order gravity (and also in supergravity, at least in its minimal formulation) are: a part quadratic in the curvature giving the propagator a UV behaviour such that only one-loop divergences can survive, and another part contributing only to vertices whose couplings can be chosen to kill the one-loop divergences. The question arises as to what the consequences of such terms on observable quantities are. In particular, it was recently shown that in the case of Bosonic super-renormalizable or finite gravitational theories \cite{amplitudes} containing terms at least quadratic in the Ricci tensor or the Ricci scalar (apart from the usual Einstein-Hilbert action operator) 
all the $n$-graviton scattering amplitudes coincide with the Einsteon-Hilbert ones. 
%{\color{red} although playing a fundamental role in determining the on-shell $n$-graviton scattering amplitudes.}
The astonishing result relied on a theorem reported in 
\cite{Anselmi:2006yh, Anselmi:2002ge, AnselmiQG} and stating that an action
\be
S^\prime (\phi_i)=S(\phi_i)+S_i F_{ij}S_j\,,
\ee
where $S_i\equiv\delta S/\delta\phi_i$ are the equations of motions for the action $S(\phi_i)$ and $F_{ij}$ is a symmetric operator generically dependent on derivatives in a local or wekly nonlocal way, can actually be recast as
\be
S^\prime (\phi_i)=S(\phi_i^\prime )\,,
\ee
by a field redefinition
\be
\phi_i^\prime=\phi_i+\Delta_{ij}\phi_j\,,
\ee
with $\Delta_{ij}$ itself  a symmetric operator dependent on derivatives. In the case under consideration we do not need to explicitly evaluate the four or n-points scattering amplitudes, but we just have to apply the field redefinition theorem to a large class of local or weakly nonlocal supergravity theories, and 
%A their
%
the outcome of any scattering amplitude can be read out of the following theorem. 

{{\bf Theorem.}} {\em All the tree-level $n$-point functions in any $N=1$ supergravity theory 
$($in particular super-renormalizable or finite$)$, with an action} 
\be 
S_{\rm NL} \! = \! \int \! d^8 z E^{-1} \left(-3 \kappa^{-2}
+ {R} \, \gamma_{{R}}(\Delta_{{R}} ) \, \bar{{R}} 
+ G^{\alpha\dot\alpha} \,  \gamma_{G}( \Delta_{G} ) \, G_{\alpha\dot\alpha} \right) % \nonumber \\
%&& \hspace{0.6cm}
%+ \int \! d^6 z  \hat{\phi}^3 
%W_{\alpha \beta \gamma} \, \gamma_{W}( \Delta_{W} )\, W^{\alpha \beta \gamma} + {\rm h.c}  
+ \int E^{-1}{\bf V}( { R}, G_{\alpha \dot{\alpha}}, W^{\alpha \beta \gamma}) \, ,
%\label{SNL}
\label{gravitySNL}
\ee
{\em can be equivalently derived from the Einstein-Hilbert Supergravity theory, 
$S_{\rm SR} \! = \! \int \! d^8 z E^{-1} \left(-3 \kappa^{-2}\right)$, provided that the potential ${\bf V}$ 
is at least quadratic in ${R}$ and/or $G_{\alpha \dot{\alpha}}$. 
In particular for any theory in which we can recast the potential in the following form 
\be
%&&
 \hspace{-0.4cm}
{\bf V} = {R} \, {\bf V_1}({ R}, G, W) 
{ R} + 
{G_{\alpha \dot{\alpha}}} {\bf V_2}( { R}, G, W)^{\alpha \dot{\alpha} \beta \dot{\beta}}
{G_{\beta \dot{\beta}}} %\nonumber \\&&
= %\overbrace{
\left( { R} , {G_{\alpha \dot{\alpha}}} \right)_{i}
%}^{E_i}
 \, {\bf V}^{ij} \, ( { R} , {G_{\beta \dot{\beta}}} )_j \equiv E_i {\bf V}^{ij} E_j , 
\label{VRic}
\ee
the theorem is valid}.

{{\bf Proof.}} 
The proof is based on the field redefinition theorem proved in \cite{Anselmi:2006yh, Anselmi:2002ge, AnselmiQG}
at perturbative level and to all orders in the Taylor expansion of the redefinition of the metric field.

We assume that we have two generally weakly nonlocal action functionals $S'(H)$  and $S(H')$, respectively defined in terms of the superfields 
$H$ and $H'$, such that
%we first modify $S'$ into 
\be
S'(H) = 
 S(H) + E_i(H) F_{i j}(H)  E_j (H) \, ,
 \label{AnselmiC}
\ee
where $F$ can contain derivative operators and ${E_i = \delta S/\delta H_i}$ are the EOM of the theory with action $S(H)$. 
The statement of the theorem is that there exists a superfield redefinition 
\be
H_i'  = H_i + \Delta_{i j} E_j  \quad \Delta_{i j} = \Delta_{j\hspace{0.03cm} i}, 
\ee
such that, perturbatively in $F$, but to all orders in powers of $F$, we have the following equivalence,
\be
S'(g) = 
S(g')  \, .
\label{FR}
%= S(g) + E_i F^{i j}  E_j (g),
\ee
In the above formula $\Delta_{ij}$ is a possibly nonlocal operator acting linearly on the EOM $E_j$, with indices $i$ and $j$ in the field space, and it is defined perturbatively 
in powers of the operator $F_{ij}(H)$, namely 
\be
 \Delta_{ij}=F_{ij}(H)+\ldots \, .
\ee
% starting at, or infinitesimally
Let us consider the 
%It is easy to prove the theorem at the 
first order in the Taylor expansion for the functional $S(H')$, which reads
\be
S(H') = S(H + \Delta H) \approx S(H) + \hspace{-0.05cm}  %\overbrace{ 
\frac{\delta S}{\delta H_i} %}^{ {\rm EOM} }
\hspace{-0.05cm}   \delta g_i = 
 S(H) + E_i   \, \delta H_i \, .
\ee
If we can find a weakly nonlocal expression for $\delta g_i$ such that 
\be
S'(H) = %S(g') \approx 
S(H) + E_i   \, \delta H_i 
\ee
(note that the arguments of the functionals $S'$ and $S$ are now the same), 
then there exists a field  redefinition $H\rightarrow H'$ satisfying \eqref{FR}. Hence, the two actions $S'(H)$ and $S(H')$ are tree-level equivalent.

As it is obvious from above, in the proof of our theorem it was crucial to use the classical EOM $E_i$. In the theory \eqref{gravitySNL} this implies ${\mathcal R} = G_{\alpha \dot{\alpha}}=0$ in vacuum.

Now we can explicitly apply the above field redefinition theorem to our class of theories \eqref{gravitySNL}, where we did not include terms with super-Weyl tensors. Since we are interested 
in $S(H') \equiv S_{\rm SG}(H')$ 
and $S'(H) \equiv S_{\rm NL}(H)$, the relation \eqref{AnselmiC} reads
\be
\hspace{-0.5cm}
S(H') \equiv 
S_{\rm SG}(H') %\equiv
 %S(H') 
 = S_{\rm SG} (H) + \left( {R} \, {\bf F_1}  {R} \right) (H) + 
 \left( G_{\alpha \dot{\alpha}} \, {\bf F_2}^{\alpha \dot{\alpha} \beta \dot{\beta}}   G_{\beta \dot{\beta}} \right) (H)
= S'(H) 
\equiv S_{\rm NL}(H) 
\label{F12}
\, .
\ee
The explicit form of ${\bf F_1}$ and ${\bf F_2}$ can be derived by comparing (\ref{F12}) and (\ref{gravitySNL}).

\section{Spacetime singularities}
In this section we discuss some aspects of the spacetime singularities in nonlocal 
supergravity. In particular we restrict our analysis to the solutions of the exact equations of motion (EOM) $E_{\alpha \dot{\alpha}} =0$ such that  ${R} = G_{\alpha \dot{\alpha}} =0$. The class of actions (\ref{SNL}) is in fact such that the EOM have the form
\be
&{ R}+(\dots) { R}=0\,,\\
&G_{\alpha \dot{\alpha}}+(\dots) G_{\alpha \dot{\alpha}}=0\,,
\ee
where  $(\dots){ R}$ and $(\dots) G_{\alpha \dot{\alpha}}$ are determined by the part quadratic in the supercurvatures. As the EOM turn out to be at least linear in $R$ and $G_{\alpha \dot{\alpha}}$ we find 

\begin{comment}
 What we need are the EOM for the theory (\ref{SNL}). However, to infer about 
spacetime singularities it is sufficient to consider the theory for $\gamma_W=0$ and to observe that 
if the EOM for the local supergravity are satisfied, then the EOM for the nonlocal supergravity are satisfied too. 
Indeed, the variation of the first operator gives ${ R}$ and $G_{\alpha \dot{\alpha}}$, while the variation of the second and third operator gives $(\dots) { R}$ and 
%and the variation of the third one produces 
 $(\dots) G_{\alpha \dot{\alpha}}$ respectively. We do not need to take care of the extra operators shortly denoted by
 $(\dots)$ because we are not interested in the full EOM.
 If we denote the exact EOM for the nonlocal supergravity by $E_{\alpha \dot{\alpha}} =0$, 
 what we just proved is the following statement,
 \end{comment}
 
 \be
 {R} = G_{\alpha \dot{\alpha}} =0 \quad \Longrightarrow \quad E_{\alpha \dot{\alpha}} =0. 
 \ee
Therefore, all solutions of the local Einstein supergravity in the vacuum are exact solutions of the nonlocal supergravity as well.  In particular the supersymmetric version of Schwarzschild metric \cite{super-sch} is still a solution in the vacuum and, although endowed with some extra features  with respect 
to its Bosonic counterpart, is still singular in $r=0$. 

In \cite{Yaodong} it was proven that Ricci-flat spacetimes are exact solutions in a large range of 
consistent theories at quantum level. In particular, singular spacetimes, like the one described by the Schwarzschild and Kerr metrics, are exact solutions. 
Therefore, spacetime singularities are still present in well defined quantum gravitational theories, and
as long as we have proved in this section local supersymmetry does not improve the things. 
However, one can still require that in a nonlocal theory the usual point-like source 
is delocalized by the form factor so that the Schwarzschild solution gets corrections at short distances and the singularity is smeared out \cite{ModestoMoffatNico, modesto}. 
%Therefore, %we observe that in 
In a nonlocal supergravity we can have singular spacetimes also in the presence of other dynamical fields then the graviton, namely the gravitino field which is its supersymmtric counterpart.
However, the regularity of the graviton and gravitino potentials %that can be easy derived 
suggests that the same delocalization mechanism proposed for the Bosonic theory \cite{ModestoMoffatNico, modesto, BM} may remove the singularities also in the supersymmetric one.
Moreover, it has been suggested that in a conformal invariant extension of the Bosonic theory 
the singularities are just artifacts of the conformal frame \cite{thooft, BarsTurok, Bars2}. 
In the same way we expect that an extension of the nonlocal supergravity to a nonlocal
conformal supergravity will heal the singularities of the theory here presented.

\section{Conclusions}
In this paper we explicitly constructed the supersymmetric extension of a class of
nonlocal super-renormalizable or finite gravitational theories \cite{Tombo, modesto} in the superspace formalism. 
Tree-level unitarity has been proved considering the linearized superfield equations of motion. 
%Expanding the gravitational 
The components' quadratic action  %in component fields the linearized action
has also been worked out and it turned out to have the same structure as the one for Einstein-Hilbert supergravity except for an overall multiplicative modification of the kinetic term by % times 
$\exp {\rm H}(\Delta)$ (where ${\rm H}$ is an entire function of the d'Alembertian operator.)
Therefore, the perturbative spectrum is the same of the local $N=1$ supergravity and it consist on the graviton and the gravitino field as well as the usual non-propagating auxiliary fields characteristic of the old-minimal formulation. 
As a simple generalization we also proposed a supersymmetric version of the Bosonic 
theory published in \cite{Briscese:2013lna}, where a super-renormalizable completion of 
the Starobinsky theory was proposed. The Bosonic spectrum of the theory in \cite{Briscese:2013lna} is enlarged to include Starobinsky's scalaron,
while the super-symmetric theory shows up four real scalars and their femionic partners. 
As an application of our nonlocal construction in superspace, we also proposed two alternative 
theories of nonlocal massive supergravity. However, these theories are not weakly nonlocal and 
their quantum properties are currently far from being clear. 
%%%
On the other hand, the weakly nonlocal supergravity (minimal or the Starobinsky one)
%At the quantum level the theory is 
are power-counting super-renormlizable and
only one loop divergences survive. However, we cannot exclude that the simpler 
theory here proposed is actually finite even at one loop, but we need and explicit computation of the beta functions. If it is not the case, two more local operators
in the superspace, both quartic in the curvatures, can do the job and make the theory completely finite 
in dimensional regularization scheme similarly to what has been recently achieved in the purely Bosonic case \cite{modestoLeslaw}. 
Moreover, in the cut-off regularization scheme one more 
divergence appears proportional to the Einstein-Hilbert supergravity operator. %three one-loop beta function can be non zero
%, but tho out of the three 
%%out of the four one-loop beta functions 
%can be make vanish at any energy scale, by just adding two extra local operators in the super-space. 
However, we can likely make zero the one-loop beta function 
for the Newton constant
just adding one extra local operator to the action 
and the theory turns out to be completely finite in the cut-off regularization scheme too. 

We also showed, generalizing a field redefinition theorem proved in 
\cite{Anselmi:2006yh, Anselmi:2002ge}, that all tree-level particle scattering amplitudes are identical to the one of local Einstein supergravity. Namely, the theory is actually local at classical perturbative level and all order in the Taylor expansion of the superfield. 

Finally, we pointed out that the kind of weak non locality needed to attain a unitary spectrum 
and finiteness
is not sufficient to sweep away the spacetime singularities %even if it is till possible to accommodate a solution to the Big-Bang or black hole 
%singularity issue  
\cite{Yaodong}.  
As a particular case, when the super-Weyl square nonlocal term is absent, the singular super Schwarzschild metric derived in 
\cite{super-sch} is an exact solution of the nonlocal supergravity (\ref{SNL}) too.  
Therefore, non locality, quantum finiteness, and super-symmetry can altogether give 
a finite theory of quantum gravity, but fail in removing the spacetime singularities. 
We suggest that only in a ``more symmetric theory", like a super-conformal invariant extension of the theory here presented, the classical spacetime singularities may be definitively wiped out. 
 An intriguing related possibility is that our finite quantum (super-)gravity is the spontaneously broken phase of a conformal invariant (super-)gravity theory \cite{preparation}. 
%Here we are assuming that, for example, Ricci flat spacetimes are exact 
%solutions of the full quantum action. 
However, it is still possible that quantum gravity can not solve completely the singularity problem of general relativity.

\acknowledgments
We would like to thank Les\l{}aw Rachwa\l{} for the many discussions and suggestions during the completion of this work.

\end{document}